\documentclass[aps,prb,amsmath,twocolumn,amssymb,titlepage,superscriptaddress,showpacs]{revtex4}
\usepackage[english]{babel}
\usepackage{graphicx}
\usepackage{transparent}
\usepackage{amsmath}
\usepackage{amssymb}
\usepackage{epstopdf}
\usepackage{amsfonts}
\usepackage{bm}
\usepackage{mathdots}
\usepackage{times}
\usepackage{pdfpages}
\usepackage{mathtools}
\usepackage{stmaryrd}
\usepackage{tabularx}
\usepackage{hhline}

\newcommand{\I}{\mathrm{i}}

\newcommand{\beq}[1]{\begin{equation} #1 \end{equation}}
\newcommand{\bsplit}[1]{\begin{equation} \begin{split} #1 \end{split} \end{equation}}

\newcommand{\astcycl}{\mathrlap{\kern0.085em{\circlearrowright}}\ast}
\newcommand{\taucycl}{\mathrlap{\kern0.42em{\bullet}}\circlearrowright}

\begin{document}
\title{Photo-induced gap closure in an excitonic insulator}
\author{Denis Gole\v z}
\affiliation{Department of Physics, University of Fribourg, 1700 Fribourg, Switzerland}
\author{Philipp Werner}
\affiliation{Department of Physics, University of Fribourg, 1700 Fribourg, Switzerland}
\author{Martin Eckstein}
\affiliation{Max Planck Research Department for Structural Dynamics, University of Hamburg-CFEL, 22761 Hamburg, Germany}


\begin{abstract}
We study the dynamical phase transition out of an excitonic insulator phase after photo-excitation using a time-dependent extension of the selfconsistent GW method. We connect the evolution of the photoemission spectra to the dynamics of the excitonic order parameter and identify two dynamical phase transition points marked by a slowdown in the relaxation: one critical point is connected with the trapping in a nonthermal state with reduced exciton density and the second corresponds to the thermal phase transition. The transfer of kinetic energy from the photoexcited carriers to the exciton condensate is shown to be the main mechanism for the gap melting. 
We analyze the low energy dynamics of screening, which strongly depends on the presence of the excitonic gap, and argue that it is difficult to interpret the static component of the screened interaction as the effective interaction of some low energy model. Instead we propose a phenomenological measure for the effective interaction which indicates that screening has minor effects on the low energy dynamics.
\end{abstract}
\maketitle

\section{Introduction}

Recent progress in time-resolved techniques has opened a new window to probe the dynamics of complex materials on the femtosecond timescale.\cite{giannetti2016} An intriguing goal of that research is to understand and possibly control the ultrafast dynamics of symmetry-broken phases such as charge density waves (CDW) or superconductivity.\cite{fausti2011} Many ordering phenomena involve an interplay between the electrons and the lattice, which provides a major challenge for the  theoretical and experimental understanding, and also links the relevant timescales to the lattice motion. In contrast, the phenomenon of exciton condensation is a symmetry breaking mechanism of purely electronic origin, which would therefore provide an ideal testbed to study the dynamics of long-range order on the electronic timescale, where the lattice can be considered as frozen. 

Excitonic phases were proposed in a seminal work \cite{mott1961} by Mott as a new insulating phase of matter in semi-metals, and later on also discussed for semiconducting materials.\cite{knox1963}  The proposed mechanism is that the Coulomb interaction binds conduction band electrons and valence band holes into pairs called excitons.  If the binding energy of such a pair is larger than the gap the pairing leads to a phase coherent state - the excitonic insulator (EI) - by a mechanism similar  to that of BCS superconductors .\cite{jerome1967} If the binding takes place between electrons and holes at different locations in the Brillouin zone, the condensate corresponds to a CDW.\cite{kohn1967}

Even though several candidate materials were proposed that may exhibits an EI phase, the mechanism is still debated in most cases. \cite{disalvo2976,kidd2002, rossnagel2002, monney2009, pillo2000} A prominent example is 1T-TiSe$_2$, which is a semi-metallic material exhibiting a commensurate CDW accompanied by a periodic lattice distortion. The simultaneous ordering of different degrees of freedom also allows for alternative explanations for the symmetry-breaking based on electron-phonon coupling,\cite{hughes1977}  which can possibly cooperate with the excitonic pairing to stabilizes the CDW,\cite{ kaneko2015, kaneko2013, zenker2014} and in addition the CDW may also compete with excitonic spin-density wave phases.\cite{zocher2011, kaneko2012, kaneko2015} An analysis of the {\em dynamics} of CDW phases after photoexcitation, and in particular the timescale for its melting, can provide a different view on the origin of the long-range order and help to disentangle various contributions. Ultra-fast melting of the CDW gap in 1T-TiSe$_2$ has been interpreted in favor of an electronic mechanism,\cite{mohr2011,rohwer2011} though also phononic contributions to the CDW  can be identified.\cite{porer2014} 

Theoretically, understanding the dynamics of symmetry broken states is a challenging problem even in a purely electronic system. One possible mechanism for the melting of the CDW in 1T-TiSe$_2$ is a photo-induced screening of the Coulomb interaction responsible for the electron-hole binding.\cite{rohwer2011} According to previous studies on semiconductors, the timescale of the buildup of screening is the inverse plasma frequency.\cite{huber2001, rossi2002, sayed1994, banyai1998} On the other hand, excitons can break up by inelastic scattering with photo-induced carriers. Naively one might expect that this process leads  to a rapid thermalization to a hot electron temperature higher than the critical temperature. However, if the symmetry breaking is described by an effective mean-field Hamiltonian, like in BCS theory, the dynamics is known to be collisionless, i.e, the melting is a dephasing of the modes.\cite{barankov2006} Investigations of interaction quenches revealed non-thermal symmetry-broken states which persist at excitation energies corresponding to effective temperatures much above the thermal critical temperatures, while the threshold behavior is characterized by critical scaling.\cite{sciolla2013,tsuji2013,werner2012}  Remarkably, this behavior can be robust beyond mean field, i.e. transient nonthermal states and the critical behavior related to a nonthermal fixed point emerge before the system is eventually attracted to the thermal fixed point.\cite{Berges2008,tsuji2013,nowak2011}

Understanding the mechanism of ultra-fast melting in an electronic symmetry-broken phase is thus a highly nontrivial task and it is important to elucidate the processes involved in order to properly interpret their experimental signatures. In this paper, we study the dynamics of an excitonic CDW insulator after photo-excitation within a two-band model. In order to capture the effects of exciton formation, incoherent electron-electron scattering, and screening in the nonequilibrium domain on the same footing, we implement the self-consistent GW approximation on the Kadanoff-Baym contour\cite{stefanucci2013} and allow for excitonic and spatial symmetry breaking.  We show that by exciting the system we can induce a transition from the excitonic insulator to a non-thermal trapped state with reduced excitonic binding or to a thermal phase without symmetry breaking, similar to the case of an antiferromagnetic insulator. \cite{tsuji2013}

This paper is organized as follows. In Sec.~\ref{Sec.:ModelMethod} we introduce the model and describe the time-dependent mean field approximation and GW approximation. In Sec.~\ref{Sec.:Results} we first introduce a simple equilibrium picture of the excitonic insulator using the mean-field approximation. Later we employ the fully self-consistent GW description  and  analyze the dynamics of the excitonic order parameter after a photodoping pulse and the gap closure in the photo-emission spectra (PES). 
In Sec.~\ref{Sec.:Discussion},  we demonstrate that also in the nonequilibrium situation the gap in the PES and the value of the order parameter are strongly correlated as one would expect for the equilibrium systems. We further discuss how one can interpret the evolution of the PES in the light of the previously discussed phenomena. 
Section~\ref{Sec.:Conclusion} is a conclusion and an outlook.

\section{Model and method}\label{Sec.:ModelMethod}

\subsection{Hamiltonian}

We study a one-dimensional two-band system of spinless Fermions described by the Hamiltonian
\begin{align}
H 
=
H_\text{kin}
+
H_\text{loc}
+
H_\text{dip}(t)
+
H_\text{int}.
\label{THEmodel}
\end{align}
The noninteracting band-electron contribution is given by
\bsplit{
H_\text{kin} + H_\text{loc}=
\sum_{k\alpha}
(\epsilon_{k\alpha}+\Delta_\alpha)
c_{k,\alpha}^\dagger c_{k,\alpha},
}
with $\epsilon_{k\alpha} = -2J_\alpha\cos(k)$ the band dispersion and $c_{k,\alpha}$ the annihilation operator for an electron with momentum $k$ in orbital $\alpha=1,2$. The crystal field splittings $\Delta_{1,2}$ are chosen such that band $1$ is occupied, except for a hole pocket close to ($k=\pi\equiv Q$), while for band $2$, only an electron pocket near $k=0$ is occupied. The second term allows to make direct (dipole allowed) inter-band transitions 
\begin{align}
H_\text{dip}(t)
=
\sum_{k}
\Big(
A(t) 
c_{k,1}^\dagger
c_{k,0}
+
\text{H.c.}
\Big),\label{Eq.:Dip}
\end{align}
to describe the laser induced population transfer between the bands. Finally we consider a density-density Coulomb interaction of the form
\begin{align}
H_\text{int} 
&=
\frac{1}{2}
\sum_{i,j}
\sum_{\alpha\alpha'}
V^{\alpha\alpha'}_{|i-j|}
n_{i\alpha}n_{j,\alpha'}
\end{align}
between lattice sites $i$ and $j$. We adopt an exponentially decaying interaction
\begin{align}
V_{l}^{\alpha,\alpha'}=U(1-\delta_{\alpha,\alpha'})\delta_{l,0} + V e^{-\lambda (|l|-1)}(1-\delta_{l,0}),
\end{align}
where we separated the local $U$ and the nonlocal part of the interaction $V$ with a characteristic screening length $1/\lambda$.
The diagonal term $U^{\alpha,\alpha}_{0}=0$ is assumed to vanish (such terms are absorbed into the single-particle part of the Hamiltonian). Throughout this paper we focus on the half-filled (particle-hole symmetric) case; this is ensured by choosing the bands symmetric like
$\Delta_2= \mu + \Delta/2$ and $\Delta_1= \mu - \Delta/2$ with $\mu=-[U+4Ve^\lambda/(e^\lambda+1)]/2.$ 
In momentum representation the interaction Hamiltonian reads
\begin{align}
H_\text{int} 
&=
\frac{1}{2L}
\sum_{k,k',q}
\sum_{\alpha\alpha'}
V_{\alpha\alpha'}(q)
c_{k+q,\alpha}^\dagger c_{k,\alpha}
c_{k'-q,\alpha'}^\dagger c_{k',\alpha'},
\label{hink-k}
\end{align}
with the vertex
\begin{align}
V_{\alpha\alpha'}(q)=U(1-\delta_{\alpha,\alpha'})
+
2V\frac{e^{2\lambda} \cos(q)-e^{\lambda}}{e^{2\lambda}-2e^{\lambda}\cos(q)+1}.
\end{align}

\begin{figure}[t]
\includegraphics[width=\columnwidth]{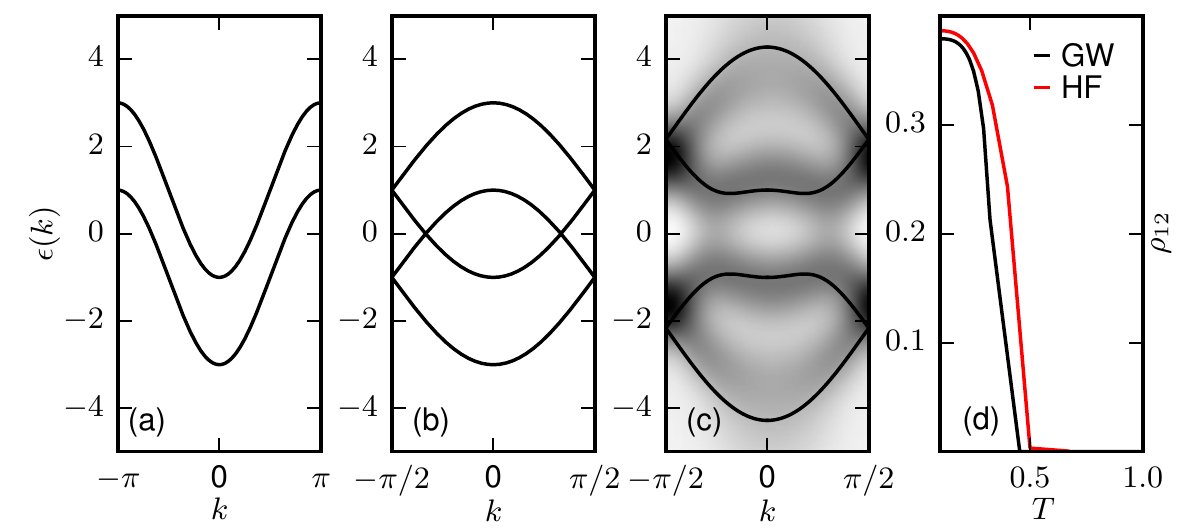}
\caption{Energy spectrum of the (a) noninteracting and (b) back-folded noninteracting problem. (c) The momentum resolved spectral function $A_{k}(\omega)$ from the GW approximation (gray). The full lines represent the energy spectrum from the self-consistent Hartree-Fock approximation. (d) Temperature dependence of the order parameter $\rho_{12}$ in the GW and Hartree-Fock (HF) approximation. All the results are for $U=3 $ and $V=1.0.$}
\label{Fig:Scheme}
\end{figure}

\subsection{Symmetry breaking}

Figure~\ref{Fig:Scheme}(a) exemplarily shows the noninteracting band structure for a band separation $\Delta=2.0$ throughout the Brillouin zone $(-\pi,\pi)$, with an electron pocket at $k=0$ and a hole pocket around $k=-\pi$. The excitonic instability arises because of the Coulomb attraction between the electrons in the lower band and holes in the higher band, leading to a condensation of the excitons which are formed across the indirect band gap.  The symmetry breaking is therefore of the type
\begin{align}
\langle c_{k+Q,1}^\dagger c_{k,2} \rangle \neq 0,
\end{align}
with $Q=\pi$, and there is a corresponding doubling of the unit cell. 

To describe the symmetry-broken phase, we back-fold the band-structure into the reduced Brillouin zone [c.f.~Fig.~\ref{Fig:Scheme}(b)]. The four bands in the reduced Brillouin zone are labelled with a super-index $\bm a \equiv(\alpha,\gamma)$ with orbital contribution $\alpha\in \{1,2\}$ and sublattice vector $\gamma\in\{0,Q\}$,
\begin{align}
\label{spinor}
\psi_k
\equiv
\begin{pmatrix}
c_{k,(1,0)}
\\
c_{k,(1,Q)}
\\
c_{k,(2,0)}
\\
c_{k,(2,Q)}
\end{pmatrix}
=
\begin{pmatrix}
c_{k,1}
\\
c_{k+Q,1}
\\
c_{k,2}
\\
c_{k+Q,2}
\end{pmatrix},
\end{align}
i.e., $\psi_{k\bm a}=c_{k+\gamma,\alpha}$. 
Below we will compute the $4\times4$ density matrix $\rho$ as
\beq{
\rho_{\bm a,\bm a'}(k)=
\langle \psi_{k\bm a'}^\dagger \psi_{k\bm a}\rangle,
\label{Ref.:density_matrix}
}
and the corresponding $4\times4$-Green-functions
\begin{align*}
\hat G_{k}(t,t')=-i
\langle
\mathcal{T}_\mathcal{C}
\psi_k(t)
\psi_k^\dagger(t')
\rangle 
\end{align*}
to characterize the order parameter and the spectral functions (photoemission spectrum).
The following subsection describes the technical details of the $4$-band Hartree-Fock and GW simulation. Results for both equilibrium and nonequilibrium states follow in Sec.~\ref{Sec.:Results}.

\subsection{Method}

Using the $4$-band representation \eqref{spinor} in the reduced Brillouin zone, the single-particle part of the Hamiltonian may be expressed as 
\begin{align}
H 
&=
H_\text{kin}
+
H_\text{loc}
+
H_\text{dip}
=
\sum_{k}
\psi_k^\dagger
\hat h(k)
\psi_k,
\end{align}
where the matrix $\hat h(k)$ is given by
\begin{align}
\begin{pmatrix}
\epsilon_{0}(k)\!+\!\Delta_{1}
&
0
&
A_k^*
&
0
\\
0
&
\epsilon_{0}(k+Q)\!+\!\Delta_{1}
&
0
&
A_{k+Q}^*
\\
A_k
&
0
&
\epsilon_{1}(k)\!+\!\Delta_{2}
&
0
\\
0
&
A_{k+Q}
&
0
&
\epsilon_{1}(k+Q)\!+\!\Delta_{2}
\end{pmatrix}
\end{align}
and the corresponding energy bands in the reduced Brillouin zone are illustrated in Fig. \ref{Fig:Scheme}(b). In the new basis we define an interaction vertex $\hat V_q$ for later use,
\begin{align}
\label{vertex}
\hat V_{(\alpha,\gamma),(\alpha',\gamma')}(q)
=
V_{\alpha,\alpha'}(q+\gamma)\,\delta_{\gamma,\gamma'}.
\end{align}

The two-band model \eqref{THEmodel} is a non-trivial many-body problem which in equilibrium has been studied by different mean-field,\cite{jerome1967,monney2009} weak-coupling\cite{bronold2006,zenker2014} and variational cluster approximations. \cite{kaneko2013,kaneko2014,kaneko2015} Here, we employ a weak-coupling method based on the random phase approximation (RPA), which is implemented on the L-shaped Kadanoff-Baym countour $\mathit{C}$ to enable a simulation of the nonequilibrium  propagation of the system, as explained in the following.
(For an introduction into the Keldysh formalism in the notation used here, see Ref.~\onlinecite{aoki2014_rev}).

\begin{figure}[t]
\includegraphics[width=\columnwidth]{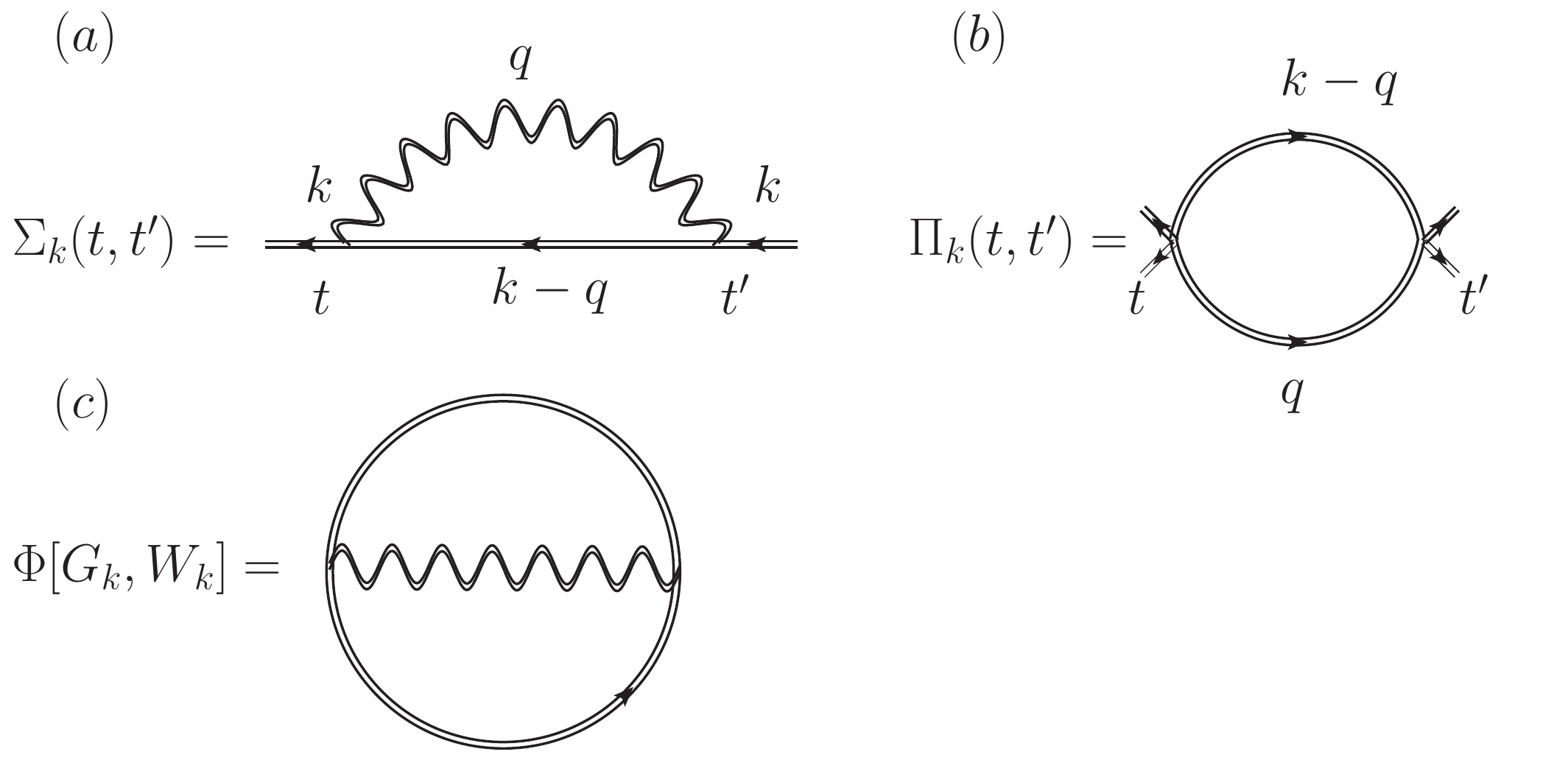}
\caption{The Feynman diagrams for the self-energy $\Sigma_k$ (a), polarization $\Pi_k$ (b) and the Luttinger-Ward functional $\Phi[G_k,W_k]$ (c).
}
\label{Fig:Diagrams}
\end{figure}

\paragraph{Hartree and Fock approximation}

Hartree and Fock self-energies are obtained by the usual mean-field decoupling of the interaction term \eqref{hink-k}.
The $4\times4$ Hartree self-energy is given by
\begin{align}
\Sigma^H_{\bm a,\bm a'}(t) =
\delta_{\alpha\alpha'}
&\sum_{\alpha_1,\gamma_1}
V^{\alpha\alpha'}(\gamma-\gamma') \nonumber \\
&\frac{1}{L}\sum_{k}{\!}^{'}
\rho_{(\alpha,\gamma+\gamma_1),(\alpha',\gamma'+\gamma_1)}(k),
\label{Eq.:Hartree}
\end{align}
where the inner sums are taken over sub-latices, bands, and 
$\sum_k'$ indicates the sum over momenta in the reduced Brillouin zone.
The $4\times4$ Fock self-energy is given by
\begin{align}
\Sigma^F_{\bm a,\bm a'}(t) = &
-\frac{1}{L} \sum_{q}{\!}^{'}
\sum_{\gamma_1} 
V^{\alpha\alpha'}(q+\gamma_1 ) \nonumber \\
&\hspace{10mm}\times \rho_{(\alpha,\gamma+\gamma_1),(\alpha',\gamma'+\gamma_1)}(k-q).
\label{Eq.:Fock}
\end{align}

\begin{figure*}[t]
\includegraphics[width=1.\linewidth]{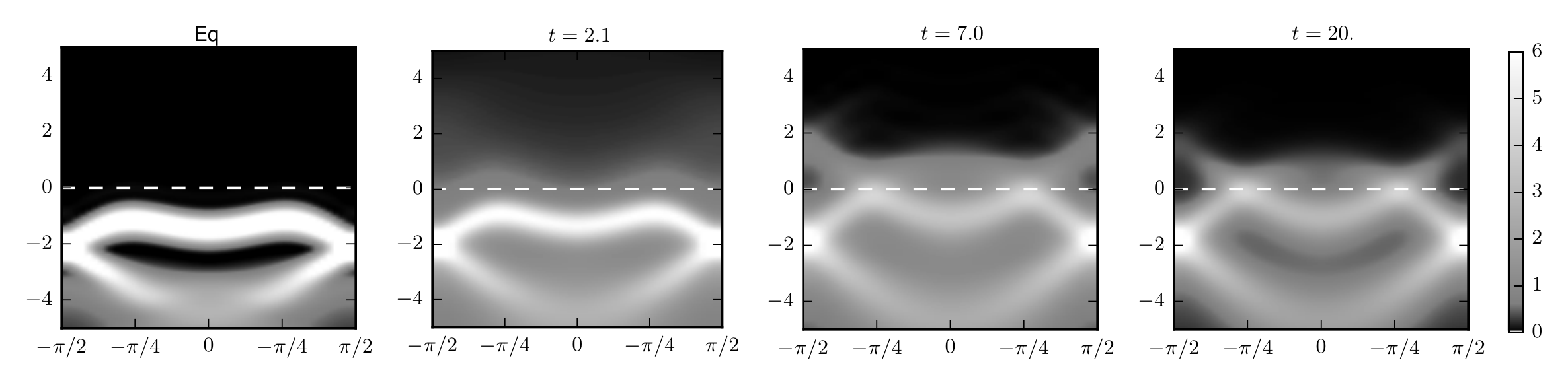}
\caption{Snapshots of the momentum resolved PES $I_k(t,\omega)$  for $U=3,V=1,\Delta=2.0,\beta=10$ after a pulse excitation with amplitude $A_0=1.4$, frequency $\omega=10.0$ and $n_p=3$ cycles. From left to the right the panels show the spectrum in  equilibrium and at  the probe times $t_{p}=2.5,7.0,20.$
}
\label{Fig:PES}
\end{figure*}

\paragraph{RPA equations}

The RPA approximation can be derived from the Luttinger-Ward functional $\Phi[G,W]$ 
(c.f.~Fig.~\ref{Fig:Diagrams})
and hence is a conserving approximation. Here we just summarize the resulting equations (for a derivation see, e.g., Ref.~\onlinecite{stefanucci2013}). The Feynmann diagrams for the electron self-energy $\Sigma_k(t,t')$ and polarization $\Pi_k(t,t')$ are shown in Figs.~\ref{Fig:Diagrams}(a,b)  and the corresponding algebraic expressions read 
\begin{align}
&\Sigma^\text{rpa}_{\bm a,\bm a'}(k,t,t')=\frac{i}{L}
\sum_{q}{}^{'}
\sum_{\gamma_1,\gamma_1'} \nonumber \\
&G_{(\alpha,\gamma-\gamma_1),(\alpha',\gamma'-\gamma_1')}(k-q;t,t')
W_{(\alpha,\gamma_1),(\alpha',\gamma_1')}(q,t,t'), \label{sigmaRPA} \\
&\Pi_{\bm a,\bm a'}(q;t,t') =\frac{-i}{L} \sum_{k}{}^{'}
\sum_{\gamma_1\gamma_1'} \nonumber \\
&G_{(\alpha,\gamma_1+\gamma),(\alpha',\gamma_1'+\gamma')}(k+q;t,t')
G_{(\alpha',\gamma_1'),(\alpha,\gamma_1)}(k;t',t).
\label{pi-k}
\end{align}
The effective interaction $\hat W$ is obtained from the integral equations
\begin{align}
\label{Eq:W-1}
\hat W_q = \hat V_q +  \hat V_q *\hat \Pi_q * \hat W_q,
\end{align}
where $*$ marks the convolution on the Kadanoff-Baym contour\cite{aoki2014_rev}.
(The bare interaction is instantaneous in time $\hat V_q(t,t')=\delta(t,t')\hat V_q$).
The Green's function, in turn, is obtained from the Dyson equation
\begin{align}
\label{dyson-g}
\hat G_k =  \hat G_k^H + \hat G_k^H*\hat \Sigma_k * \hat G_k,
\end{align}
where 
\begin{align}
\hat G_k^H = (i\partial_t - \hat h_k - \hat \Sigma^H)^{-1}
\end{align} 
is the Hartree Green's function. (Note that inserting the instantaneous contribution of $W$ into \eqref{sigmaRPA} results in the Fock-self-energy, so at the level of RPA, the full self-energy is 
$\Sigma_k(t,t') = \Sigma^\text{rpa}(t,t') + \Sigma^\text{H}(t)\delta(t,t')$.)

In the actual implementation, we treat the time-local part of $\hat W_q$ separately. We define the full susceptibility
$\hat \chi_k$ by the RPA series  $\hat \chi_k =  \hat \Pi_k + \hat \Pi_k*\hat V_k*\hat \Pi_k + \hat \Pi_k*\hat V_k *\hat \Pi_k*\hat V_k *\hat \Pi_k  + ... $,
which is numerically evaluated by solving the integral equation 
\begin{align}
\label{dyson-chi}
\hat \chi_k =\hat \Pi_k+ \hat \Pi_k*\hat V_k*\chi_k
\end{align}
on the contour ${\mathit C}$. From this we obtain $\hat W_k = \hat V_k + \hat V_k *\hat \chi_k *\hat V_k$. The solution of the integral equations \eqref{dyson-g} and \eqref{dyson-chi}
for each point $k$ in the Brillouin zone can be distributed over several compute notes. The evaluation of the momentum sums in \eqref{sigmaRPA} and \eqref{pi-k} then requires an efficient implementation of the collective communication.

\section{Results} \label{Sec.:Results}
\subsection{Equilibrium properties}

In the following section we present results for the photo-induced dynamics of the two-band model Eq.~\eqref{THEmodel}. Figure~\ref{Fig:Scheme}(a) shows the noninteracting band structure for a band separation $\Delta=2.0$ 
with an electron pocket at $k=0$ and a hole pocket around $k=-\pi$, while Fig.~\ref{Fig:Scheme}(b) shows the same data back-folded into the reduced Brillouin zone. The excitonic CDW phase can be stabilized  
for a wide range of $\Delta$. 
The symmetry breaking is evidenced by the off-diagonal components of the density matrix [c.f. Eq.~\eqref{Ref.:density_matrix}].  As an order parameter, we plot in Fig.~\ref{Fig:Scheme}(d) the temperature-dependent local (momentum-averaged) value $\rho_{12}$. (Other off-diagonal components of $\rho$ show the same behavior). As expected, the order parameter is slightly larger in the static Hartree-Fock description, compared to GW, which also includes fluctuations in the self-energy. In the symmetry-broken phase, the Hartree-Fock self-energy  leads to a shift of the energy bands and an opening of the gap, as shown by the solid lines in Fig.~\ref{Fig:Scheme}(c). The main effect of the RPA on the equilibrium band-structure is a lifetime-broadening of these bands, in particular the high-energy bands, as shown in Fig.~\ref{Fig:Scheme}(c).

\begin{figure*}[t]
\includegraphics{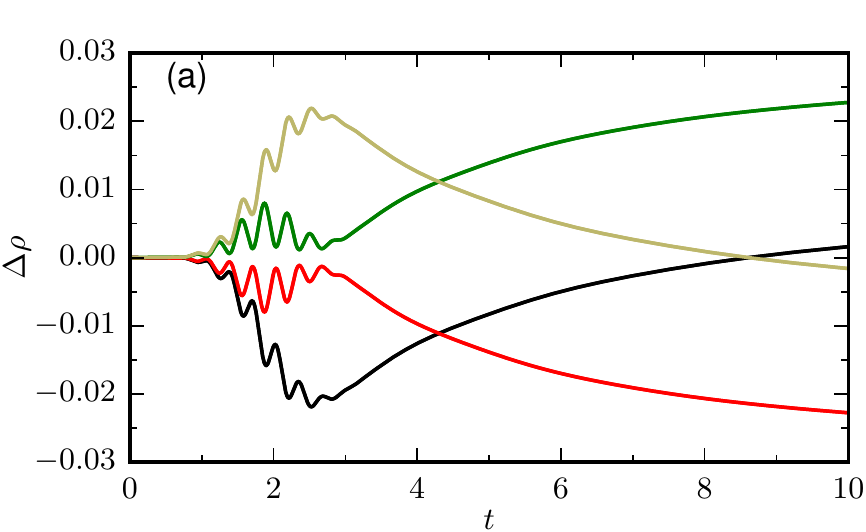}
\includegraphics{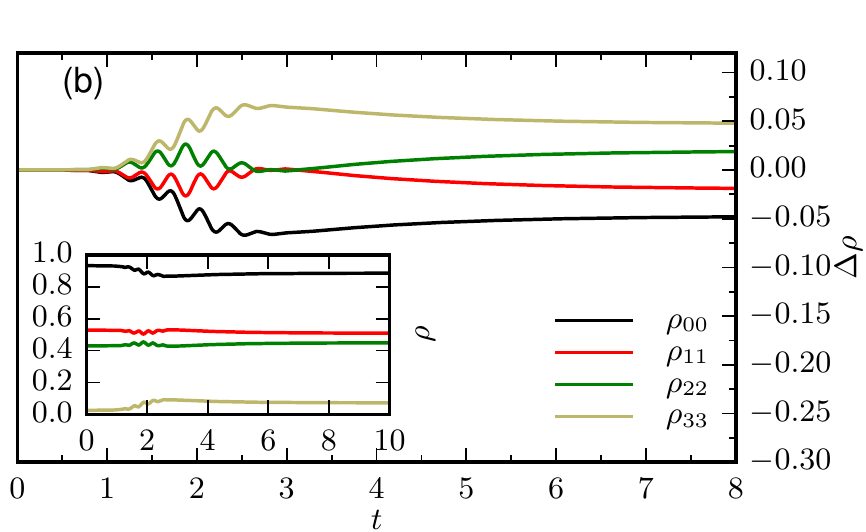} \\
\includegraphics{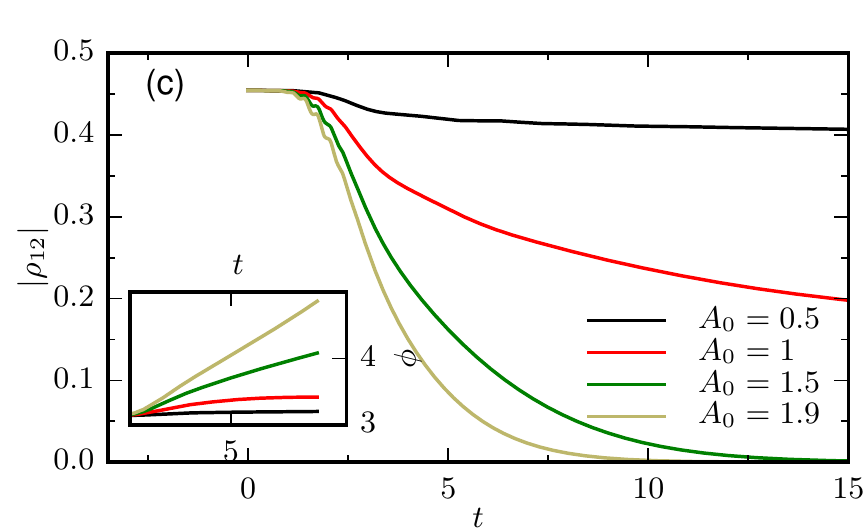}
\includegraphics{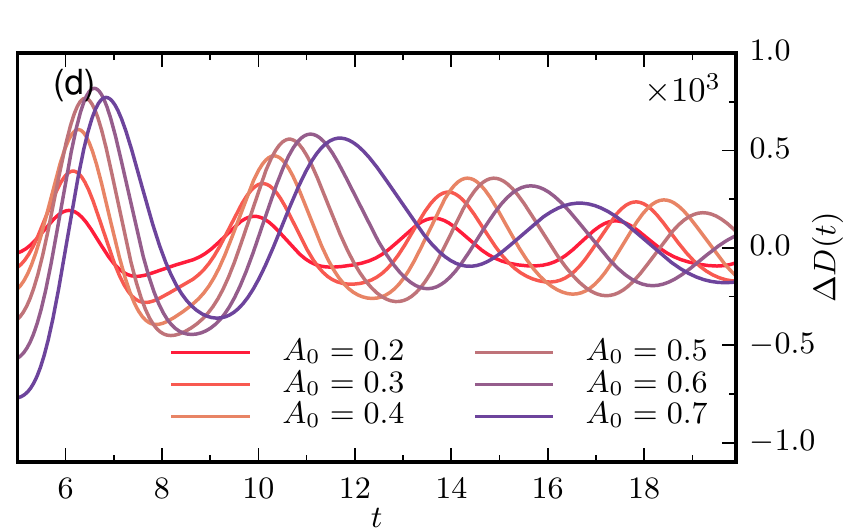}
\caption{ 
The change in the local density matrix $\Delta \rho_{ij}$ for excitation $A_0=1.1$ (a) and $A_0=2.0$ (b). The inset of (b) shows the actual values of the density matrix $\rho(t) $ for pulse strength $A_0=2.0$. The components that are not shown are small in comparison and quickly damped after the pulse. The absolute value (c) and the phase (inset) of the local  order parameter $\rho_{12}^\text{loc}$ for  pulse amplitude $A_0=0.5,1,1.5,1.9.$. The extracted amplitude modes (d) were  obtained by subtracting the exponential fit to the dynamics of the order parameter for the excitation strengths $A_0=0.2,0.3,\ldots,0.7$.}
\label{Fig.:density_matrix}
\end{figure*}

\subsection{Gap-closing in time-resolved photoemission}

We now induce the dynamics of the system by exciting the equilibrium state with a short pulse which allows direct dipolar transitions between the bands, see Eq.~\eqref{Eq.:Dip}. Unless stated otherwise, we will use the parameters of Fig.~\ref{Fig:Scheme}(c) in the subsequent simulations, which corresponds to an initial state well within the symmetry-broken region of the phase diagram, i.e., $\Delta=2.0$,  $U=3.0$, $V=1$,  inverse temperature $\beta=10$, and a rather short-ranged non-local interaction ($\lambda=5$). We use the few-cycle pump pulse $A(t)=A_0 \sin(\omega(t-t_0)) e^{-4.6(t-t_0)^2/t_0^2},$ with frequency $\omega$, amplitude $A_0$, and the width of the pulse envelope $t_0=2\pi n_p/\omega,$ where $n_p$ is the number of cycles. 

To get a first glance on the dynamics we compute the time- and angle-resolved photo-emission spectrum $I_{k}(\omega,t_p)$, which measures the photo-emitted intensity as a function of the probe time $t_p$, and of the energy $\omega$ and momentum $k$ of the photo-emitted electron.
The intensity $I_{k}(\omega,t_p)$ can be obtained from the single-particle Green's function $G_{k}(t,t')$
\begin{align}
\label{PES}
I(\omega,t_p)=&-\I \int dt dt' S(t) S(t') 
e^{\I\omega (t-t')} G_{k}^<(t_p+t,t_p+t'),
\end{align}
where $S(t)=\exp(-t^2/2\delta^2)$  is the envelope of the probe pulse with some duration $\delta$.\cite{freericks09}
Note that this expression neglects matrix element effects, which would just lead to a different intensity in various spots of the Brillouin zone, but would not influence the time-dependence of the signal. Furthermore, because photoemission is not band selective, $G_{k} \equiv \text{Tr} \,\hat G_k$ is traced over the orbital indices in Eq.~\eqref{PES}.

Snapshots of the photoemission spectrum $I(\omega,t_p)$ are plotted in Fig. \ref{Fig:PES} for the initial equilibrium state (panel labeled ``Eq") and for three times after the excitation. To resolve dynamics on the timescale of the inverse hopping, we have chosen a rather short probe duration $\delta=3$. The corresponding  uncertainty-limited frequency resolution ($1/\delta$) is nevertheless sufficient to reveal the bands in the equilibrium spectrum (compare Fig.~\ref{Fig:Scheme}(c) and Fig.~\ref{Fig:PES}(a)). Following the strong excitation, one observes a partial photo-induced population in the upper bands,
a subsequent closure of the gap, and  a restoration of the high-symmetry band-structure. For a weaker excitation, the gap would only partially close within the time window of our simulation.  Photo-emission is thus well suited  to probe the photo-induced melting of the excitonic CDW. To understand the mechanisms behind the photoinduced melting, we will now systematically analyze  the behavior of the order parameter and later return to the discussion of the photoemission spectrum.

\subsection{Nonthermal melting of the order parameter}

Figure~\ref{Fig.:density_matrix} shows the different elements of  the momentum-averaged density matrix $\rho^\text{loc}$ (Eq.~(\ref{Ref.:density_matrix})) for the two excitation strengths $A_0=1.1$ and $A_0=2.0$. The diagonal occupations $\rho_{00}$ to $\rho_{33}$ correspond to the occupation of the four (bare) bands (from bottom to top), and their time-evolution indicates the photo-induced  population shift. During the pulse $\rho_{00}$ and $\rho_{33}$ decrease and increase, respectively, which implies that the  population is predominantly transferred from the lowest to the highest band, as expected for a large pulse frequency $\omega=10.$  Subsequently, this photo-induced population relaxes to the lower bands, leading to an increase of $\rho_{22}$ and a decrease of $\rho_{11}$. We will see in the following that the energy transfer associated with the cooling of photo-carriers is related to the melting of the condensate. 

\begin{figure*}[ht]
\includegraphics{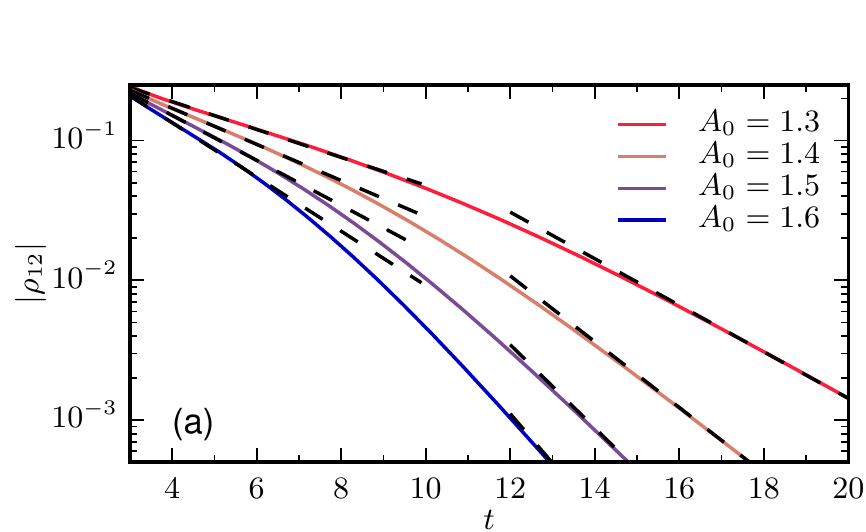}
\includegraphics{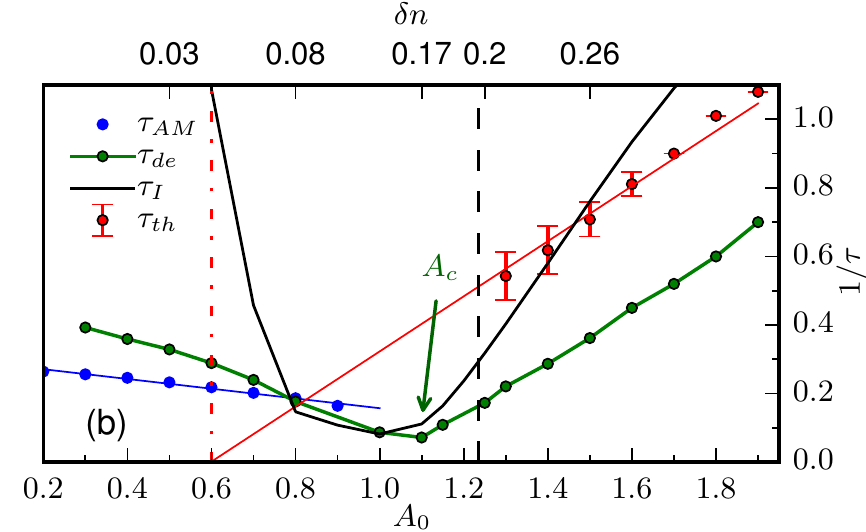}
\caption{(a) Logarithmic plot of the order parameter dynamics  $\rho_{12}$ for several excitation strengths. The dahsed straight lines show fits of the two exponential relaxations. (b) The inverse relaxation times $1/\tau$ of the initial ($\tau_\text{de}$) and long time ($\tau_\text{th}$) dynamics of the order parameter $\rho_{12}$,  the static components of the photoemission spectrum $\tau_{I}$ and the inverse oscillation times of the amplitude modes $\tau_\text{AM}$ as a function of the pulse amplitude $A_0.$ On the upper horizontal axis we indicate the amount of excited charge carriers. The red (black) dashed line corresponds to the expected point of the equilibrium phase transition due to modulation of the temperature (chemical doping). The red solid line is a guide to the eye.
}
\label{Fig.:Relax_time}
\end{figure*}

In Fig.~\ref{Fig.:density_matrix}(c) the absolute values $|\rho_{12}|$ and the phase (inset) of the order parameter are plotted  for different excitations strengths $A_0$.  After the pulse the absolute value of the offdiagonal components decreases. For weaker pulses $A_0\leq 1.1$, it remains nonzero in the long-time limit, while for the larger amplitudes it vanishes, with a critical excitation strength $A_c\approx 1.1$ for $U=3$, $V=1$. The phase of the order parameter (inset of Fig.~\ref{Fig.:density_matrix}(c)) also shows a nontrivial dynamics with a slow drift in the long-time limit in the case of a non-vanishing order parameter. On top of the overall reduction of the order parameter there are weak superimposed oscillations corresponding to the amplitude mode of the order parameter. We extract these oscillations by subtracting the slow exponential background, see Fig. \ref{Fig.:density_matrix}(d). The amplitude mode gets strongly damped and its frequency is reduced as we increase the excitation strength.

In equilibrium the transition into the high-symmetry state can be achieved by raising the temperature of the system. A naive guess might thus be that the threshold amplitude $A_c$ for the melting of the excitonic condensate is such that the total energy after the excitation corresponds to the thermal energy of the system at the equilibrium critical temperature. 
However, the final energy after the pulse $A_c$ is $E_\text{fin}=1.15$ which correspond to $\beta_\text{eff}=0.9$, while the equilibrium transition temperature is $\beta_\text{eq} \approx 2.5$. In turn, the energy corresponding to the critical temperature $\beta_\text{eff}=0.9$ is transferred to the system already by a much weaker pulse of amplitude $A_\text{th}=0.6$. This shows that the system is trapped in a transient nonthermal symmetry-broken state  and the threshold is related to the non-thermal critical point. Because the dynamics described within the GW method goes beyond the collisionless mean-field description, we expect that on longer timescales, the system evolves to the thermal state, but these times are not not accessible in our simulation for the all parameters.

To further characterize the nonthermal melting threshold in the present case, we now quantitatively analyze the relaxation times of the order parameter. For the excitation above the critical pulse amplitude, $A_0>A_c$, the order parameter  shows an exponential time-evolution with two characteristic times, the initial relaxation to the nonthermal state, which we attribute to the dephasing, and for longer times the relaxation to the thermal state, i.e., thermalization, see Fig. \ref{Fig.:Relax_time}(a). Both relaxation times exhibit a slowdown as $A_0$ is decreased and the critical point for the initial dephasing process agrees with the previously determined critical excitation strength $A_c \approx 1.1$, see Fig.~\ref{Fig.:Relax_time}(b). For an excitation strength below the critical pulse amplitude, $A_0<A_c$, the system exhibits a single exponential relaxation to the trapped state and this relaxation time  shows a slowdown at the same critical point. In this parameter region the system also shows damped amplitude oscillations, whose frequency $1/\tau_\text{AM}$ decreases with increasing pulse strength.
However the identification of the point of divergence is difficult due to the strong damping of the oscillations as one approaches the critical excitation strength $A_c$.

\subsection{Non-thermal distribution functions}

\begin{figure*}[ht]
\includegraphics{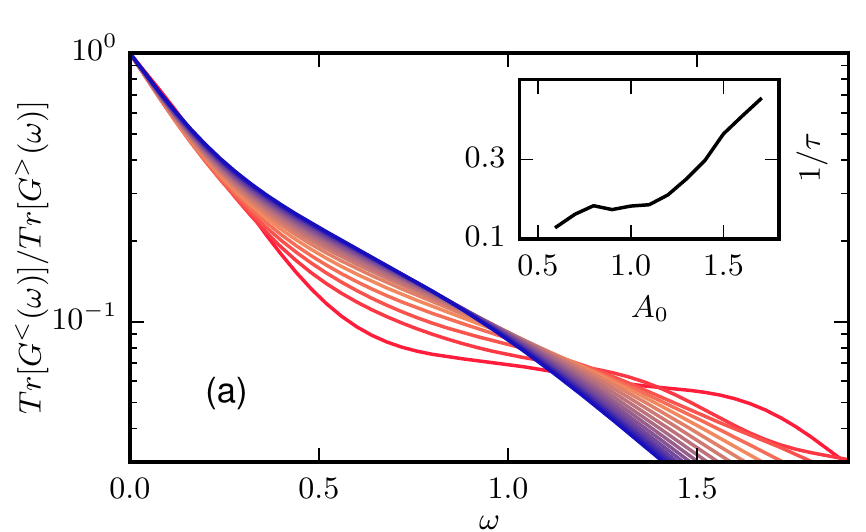}
\includegraphics{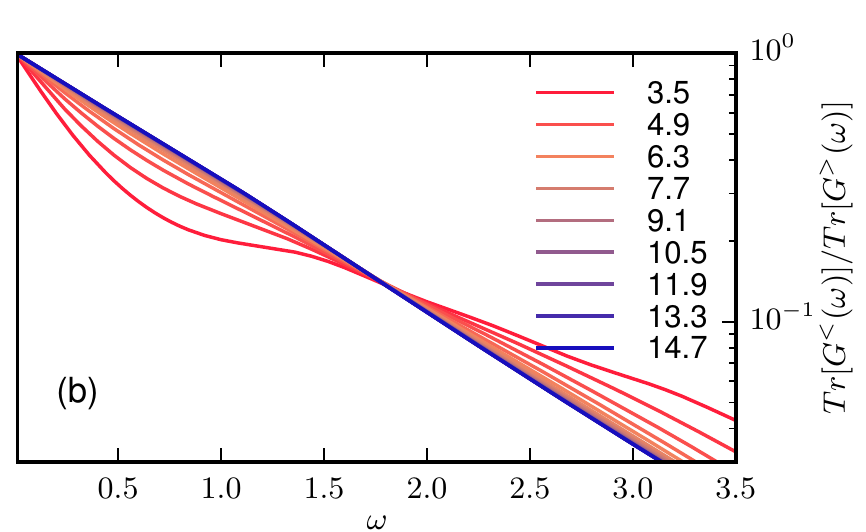}
\caption{
Time dependence of the ratio $\text{Tr}[G^<(t,\omega)]/\text{Tr}[G^>(t,\omega)]$ on logarithmic scale after an excitation with pulse strength $A_0=1.0 < A_c$ (a) and $A_0=2.0>A_c$ (b). The inset shows the relaxation time extracted from the relaxation of this ratio in the high-energy $(\omega\gtrsim 1$) region.}
\label{Fig.:therm}
\end{figure*}

The nonthermal nature of the state can also be accessed directly: In any equilibrium state, Green's functions would satisfy the fluctuation-dissipation theorem which relates spectral functions and occupation functions: $G^>(\omega)/G^<(\omega)=\exp(-\beta\omega)$. By analyzing the ratio $\text{Tr}[G^>(t,\omega)]/\text{Tr}[G^<(t,\omega)] $ in the nonequilibrium case and comparing to 
the function $\exp(-\beta_\text{eff}(t,\omega) \omega)$, we can thus see if the state in the long-time limit can be described by a thermal state at an effective temperature $\beta_\text{eff}$ 
or if a more complicated description is needed, e.g., a model with several effective temperatures. 
For an  excitation far above the critical excitation, $A_0=2.0>A_c$, this ratio in the long-time limit shows a nearly exponential $\omega$-dependence, see Fig.~\ref{Fig.:therm}(b), so the system can be described by a thermal state at elevated temperature. The long-time dynamics for excitations below the critical excitation, $A_0=1.0<A_c$, shown in Fig.~\ref{Fig.:therm}(a), however, shows a cooling of the high energy part of the spectrum ($\omega\gtrsim 1.0$). The low-energy part, on the other hand, is excited as the relaxation progresses, and exhibits a non-exponential dependence on the energy. This clearly shows that one cannot apply a single-temperature description. Interestingly, an effective temperature fit shows that for $A_0=1.0$ the very low  ($\omega \lesssim 0.3$) and high ($\omega\gtrsim 1$) energy parts have $\beta_\text{eff}\approx3.5,$ which is below the equilibrium transition temperature, while  the intermediate energy region is characterized by $\beta_\text{eff} \approx 2$, which is above the thermal transition point. 

By fitting the high-energy part of the given ratio we extracted the time dependence of the effective temperature $\beta_\text{eff}(t,\omega>1.0),$ which exhibits an exponential relaxation. The relaxation times, shown in the inset of Fig.~\ref{Fig.:therm}(a), decrease  monotonously with the pulse amplitude $A_0$, with only a weak dip at the previously determined critical excitation $A_c.$ This result is not surprising, since the relaxation time of $\beta_\text{eff}(t,\omega>1.0)$ is determined by kinetic processes at high energies, that are only weakly coupled to the low energy dynamics governed by the presence of the gap.

\subsection{Mechanism - Relation to photoexcited carrier density}

\begin{figure*}[ht]
\includegraphics{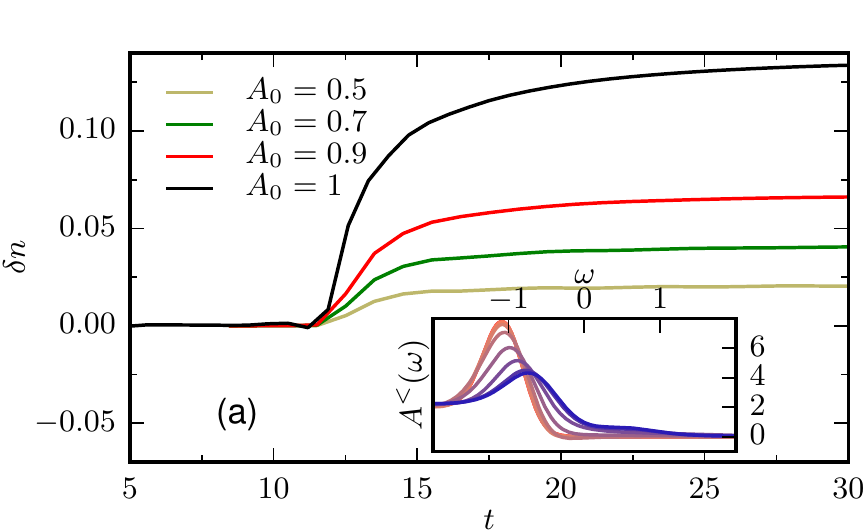} 
\includegraphics{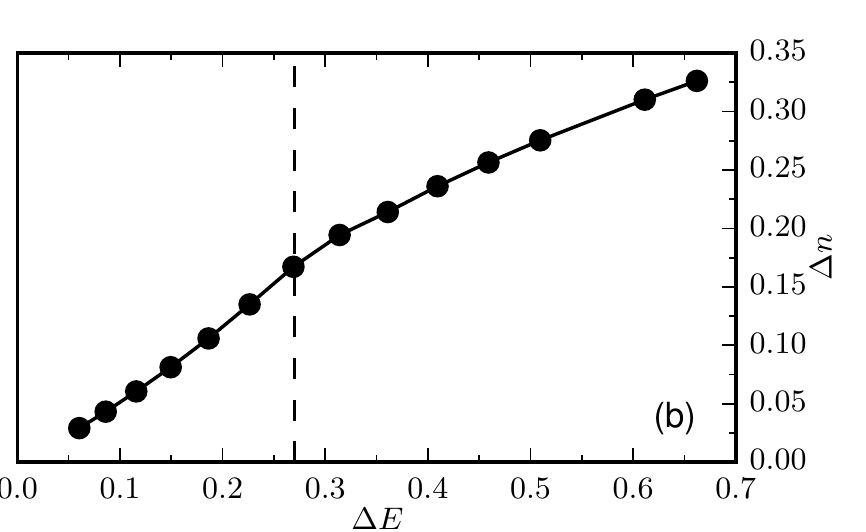}
\caption{(a) Time evolution of the excitation density $\Delta n$ for different pulse strengths $A_0$ where the center of the pulse is shifted to $t_{p}=10.5.$ The inset represent the time evolution of the occupied part of the spectrum for excitation strength $A_0=1$ with the color scale as in the previous figure. (b) Excitation density $\Delta n$ as a function of the absorbed energy $\Delta E$. The vertical dashed line correspond to the critical excitation strength $A_c.$}
\label{Fig:occupation}
\end{figure*}

By photoexciting the system we introduce additional electrons and holes and the presence of the gap prevents them from recombining. This argument suggests that the mechanism for melting of the gap is a transfer of kinetic energy of photoexited carriers to the condensate. 
High energy particle-hole pairs relax via electron-electron scattering, which breaks several bound pairs into additional free charge carriers.  
If a carrier relaxes within the upper band and breaks up an exciton, this is an impact ionization process \cite{werner2014} which results in a carrier multiplication. The process is evidenced in several ways: 

(i) The carrier production is directly seen from the time-dependent increase of the carrier number $\delta n(t)=\int_0^{\infty}d\omega [A^<(\omega,t)-A^<(\omega,0)]$ in Fig.~\ref{Fig:occupation}(a). The energy distribution of the excited charge carriers is visible in the previously introduced ratio $\text{Tr}[G^>(t,\omega)]/\text{Tr}[G^<(t,\omega)],$ see Fig.\ref{Fig.:therm}(a), with the bump (increase) in the distribution function at the energy corresponding to the excess of the charge carriers ($\omega \approx 1.0$).
Furthermore, one can look at the dependence of the number of excited carriers on the absorbed energy.
The critical excitation strength $A_c$ corresponds to an absorbed energy $\Delta E\approx 0.26$.
Above and below the critical excitation the amount of the excited charge carries indeed depends approximately linearly on the absorbed energy, see Fig.~\ref{Fig:occupation}(b). Below the critical excitation strength the ratio of the excited charge carriers and absorbed energy is larger than above. Since the condensate is only present  below the critical excitation strength the pair breaking processes results in a stronger dependence of $\Delta n$ on the absorbed energy for $A_0<A_c$. 
This non-linear dependence was actually discussed in a recent experiment,\cite{porer2014} which demonstrated two different linear dependences of the number of excited charge carriers versus absorbed energy for excitation strengths below and above the critical amplitude, though the carrier number was obtained from reflectivity data.


(ii) In the population of the bare bands, Fig.~\ref{Fig.:density_matrix}(a) and (b), the occupations of the lowest ($\rho_{00}$) and the highest ($\rho_{33}$) band relax almost to the initial values at weak excitation in the long-time limit, while a substantial population remains for strong excitation ($A_0=2.0$). Since in this case the condensate is melted, photo-induced carriers have no channel for energy relaxation left.

These arguments suggest that the state after an excitation with $A_0<A_c$ can be compared to a chemically doped equilibrium system. We would like to analyze this scenario more quantitatively by considering the amount of excited charge carriers $\delta n(t).$ As we excite the system above the critical excitation strength $A_0>A_c$ the gap is melted and the number of excited charge carriers becomes  more ill defined (see the inset of Fig.~\ref{Fig:occupation}(a) that shows the time evolution of the occupied part of the spectrum), but it  still provides a measure for the amount of excited charge carriers. 

In order to test the hypothesis that the nonthermal critical point corresponds to the equilibrium critical point obtained by doping we compare the number of the excited charge carriers in the long-time limit $n(t_\text{max})$ to the critical equilibrium value of doped holes and electrons $n_h+n_e$ at the point of the doping-induced transition, see black dashed line in Fig.~\ref{Fig.:Relax_time}(a). The comparison with this doping-induced critical point shows a better agreement than the comparison to the thermal critical point of the heated system. The equilibrium doping-induced phase transition happens close to the point where the chemical potential exceeds the top of the hole pocket so that there are no holes present to form excitonic pairs. This suggest that also the dynamical phase transition happens as a consequence of the filling-up of the hole pocket via photo-excitation.

\subsection{Screened interactions}

In this part we discuss how the screening properties are changed by the excitation, and in particular also by the destruction of the excitonic pairing.
For this purpose we compute the fully screened interaction $W_q$ [c.f.~Eq.~\eqref{Eq:W-1}], which is by definition related to the dielectric constant $\varepsilon$ and the bare interaction $V_q$ through $W_q=\varepsilon_q^{-1} V_q$. Including the fluctuations on the RPA level, retardation effects are induced in $W_q$. In nonequilibrium $W_q(t,t')$ therefore depends on two time arguments, and in order to get insights into the dynamics of screening we introduce a partial Fourier transform to define a time- and frequency-dependent interaction $W_{q}(\omega,t)=\int_t^{t+t_{max}}dt'e^{\I\omega t'} W^{R}_q(t',t).$

\begin{figure*}[ht]
\includegraphics{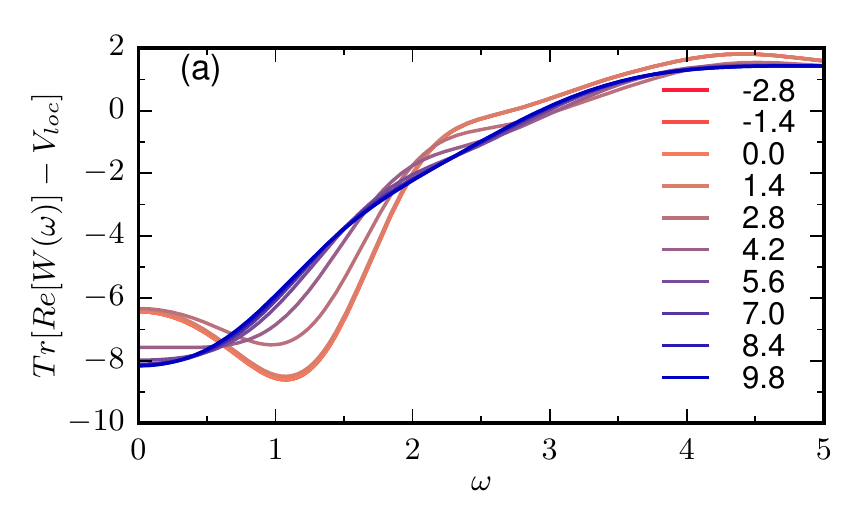}
\includegraphics{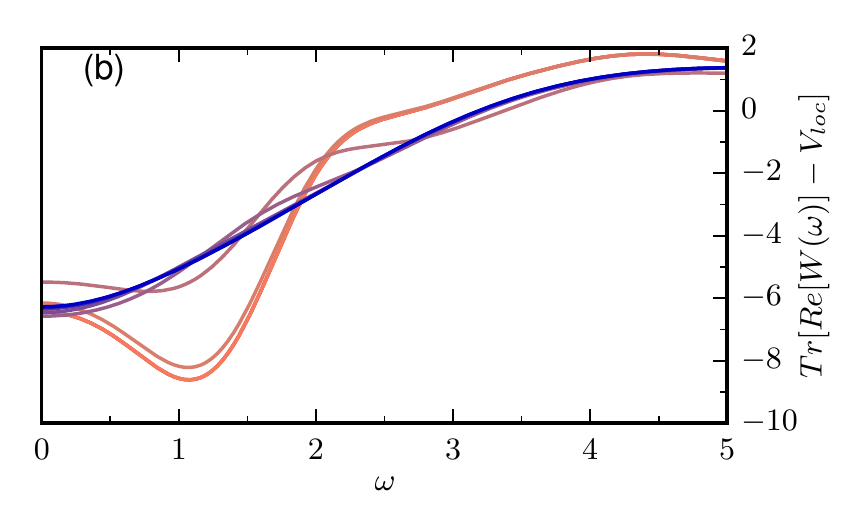}\\
\includegraphics{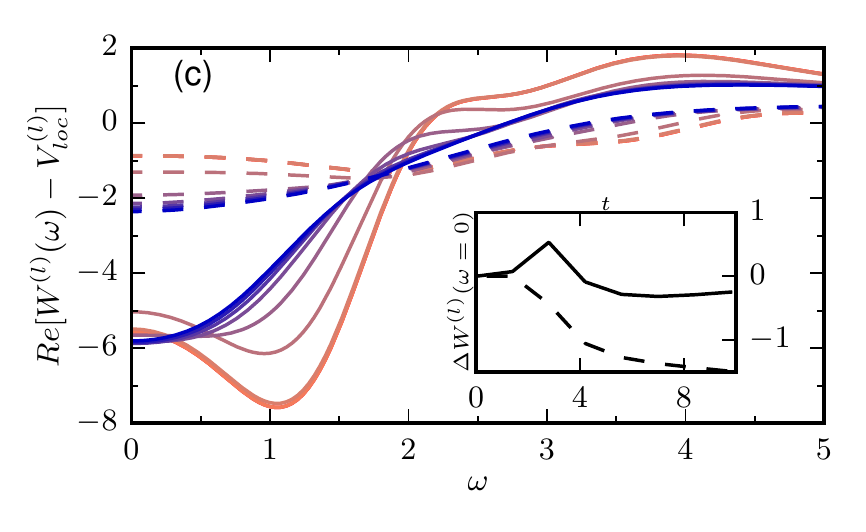}
\includegraphics{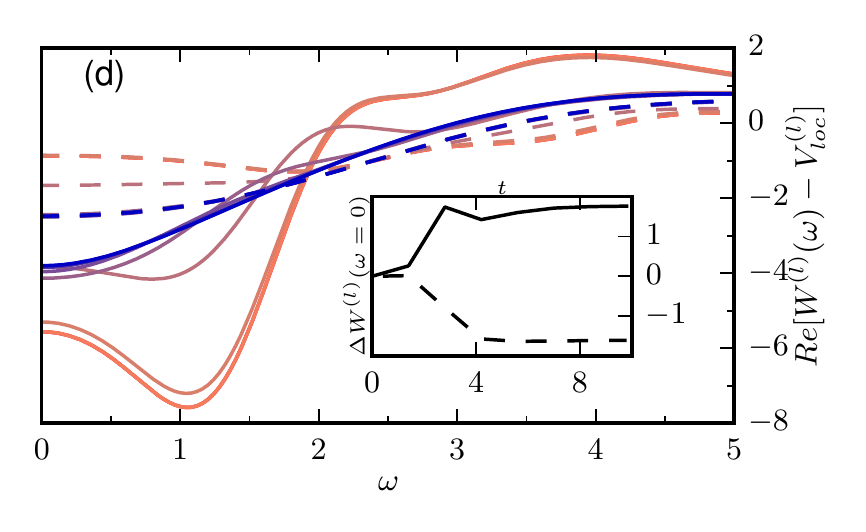}
\caption{
The trace over the real part of the partial Fourier transform of the local screened interaction 
$\text{Tr}[Re[W_{loc}](\omega,t)-U]$ for $U=3.0$ with the pulse amplitude of $A_0=1.1$ (a) and $A_0=2.0$ (b) for times before, during and after the pulse. The time $t=0$ corresponds to the center of the pulse. The dashed (full) line represent the symmetric $l=1$ (antisymmetric $l=0$) eigenmode of the effective interaction $Re[W_\text{loc}^{(l)}](\omega,t)-U$ for $A_0=1.1$ (c) and $A_0=2.0$ (d). The insets present the time evolution of the change in the static component of the eigenmodes.}
\label{Fig.:W}
\end{figure*}

For simplicity we will first analyze the local interaction traced over orbitals, $W_\text{loc}=\frac{1}{L}\sum_{q} \text{Tr} \hat W_{q}$. 
(In an experimental setup the direct measurement of $W_q$ would be related to the charge susceptibility, and thus the trace over orbitals is related to the most straightforward orbital non-selective measurement of the susceptibility.)
The equilibrium state of the system (see data in Fig.~\ref{Fig.:W} before the pulse, $t=-2.8$) shows a sizeable reduction of the effective static interaction. By exciting the system below the critical pulse amplitude, $A_0=1.0<A_c$, the static and low-energy interaction is screened, while at the energies of the order of the gap size ($0.7 \lesssim \omega \lesssim 1.7$) it exhibits a strong anti-screening effect, see Fig. \ref{Fig.:W}(a). 
For excitations above the critical amplitude, $A_0=2 >A_c$, the effective interaction after the pulse is less screened for all energies (see Fig.~\ref{Fig.:W}(b)), with the most pronounced effect at energies comparable to the gap size. 

In order to get more insight into the dynamics of screening we perform an eigenvalue decomposition of the screened interaction with respect to the orbital degrees if freedom. This yields the modes $W^{(\pm)}$ which correspond to fluctuations of the total density $n_1+n_2$ and the relative density $n_1-n_2$ of the bands $ \alpha=1,2$ in the unfolded Brillouin zone. Technically, this amounts to taking the momentum-diagonal interaction for a momentum $q+\gamma$ in the full Brillouin zone, $\tilde W^{\alpha,\alpha'}_{q+\gamma} \equiv (\hat W_{q})_{(\alpha,\gamma),(\alpha',\gamma)}$ for $\gamma\in \{0,Q\}$, which is a $2\times2$ matrix in orbital space, averaging over momenta, $\tilde W^{\alpha,\alpha'}_\text{loc} = \frac{1}{L}\sum_{q}'\sum_{\gamma} \tilde W^{\alpha,\alpha'}_{q+\gamma}$, and making the projection $ W^{\pm} = \eta_{\pm}^t \tilde W_\text{loc} \eta_{\pm} $ with the vectors $\eta_{\pm} = \begin{pmatrix}1,&\pm1\end{pmatrix}/\sqrt{2} $ onto the symmetric and antisymmetric channel.

In the case of a weak pulse, see Fig.~\ref{Fig.:W}(c), when a substantial gap is still present in the spectral function, the anti-symmetric mode $W^{(-)}_\text{loc}$ shows a strong reduction of the screening on the energy scale of the gap size, while the low energy screening has barely changed. For excitations above the critical amplitude, $A_0=2>A_c$, the anti-symmetric mode shows a strong reduction of  screening also on low energy scales, see Fig.~\ref{Fig.:W}(d). We interpret this that by photo-exciting the system we introduce additional electrons into the hole pocket, which reduces the amount of possible low-energy excitations between the bands and therefore reduces the screening. A qualitatively similar effect is present in equilibrium if we chemically dope the system. In contrast for the symmetric mode $W^{(+)}_\text{loc}$ the low-energy screening is enhanced  both by weak and strong  excitations. The increase of the low energy screening in the symmetric mode can be understood as an increase of the total amount of free charge carriers introduced into the system by photo-doping. Therefore the two eigenmodes tend to compete with each other, which leads to a nontrivial modification of the screening for the pulse strength below and above the critical amplitude $A_c$. 

These results already show that the interpretation of the static value $W(\omega=0)$ of the screened interaction as an effective interaction in a low-energy model which determines the exciton binding energy is difficult, as the interaction itself has a strong frequency dependence on the energy scale of the gap, and even for excitations close to the melting threshold there is an opposite effect of the photo-excitation on $W(\omega)$ on the energy scale of the gap and at $\omega=0$. Just by looking at the value of $W$, it is thus not easy to conclude whether the change of the screening can be viewed as a driving force for the melting of the condensate. In order to overcome this issue we will approach this question from a more phenomenological perspective in the next section.

\section{Discussion}\label{Sec.:Discussion}
\begin{figure*}[ht]
\includegraphics{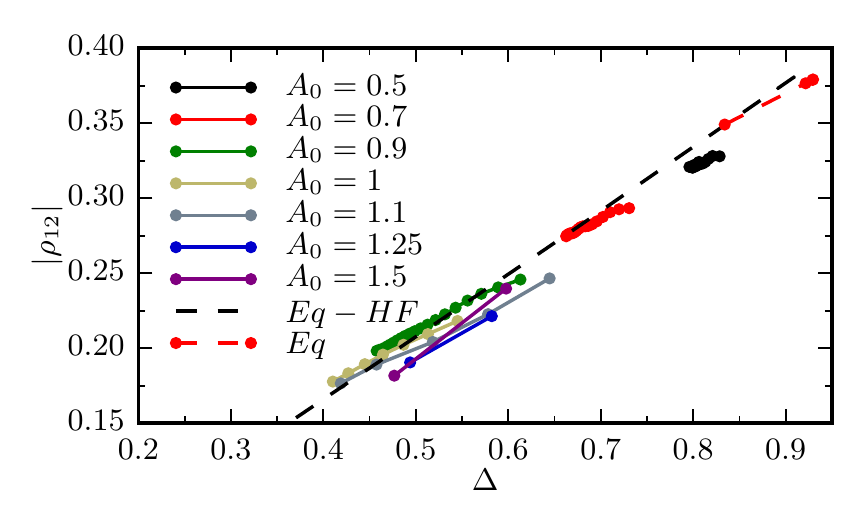}
\caption{Absolute value of the order parameter $|\rho_{12}|$ as a function of the gap size $\Delta$ determined from the photoemission spectrum after a pulse with excitation strength $A_0=0.5,\ldots,1.25$ and the equilibrium dependence for different temperatures $\beta=1,2,4,6,8,10.$ The black dashed line represents the ratio obtained from the equilibrium Hartree-Fock calculations in the  temperature range $\beta=0.2,\ldots,10.$}
\label{Fig.:RhoVsDelta}
\end{figure*}

In the light of the previous discussions we will now return to the interpretation of the photoemission spectrum. In a photoemission experiment  one can observe the dynamics of the gap, while most of the discussion above has focused on the order parameter. In a nonequilibrium situation and in the presence of dynamical screening the relation between the gap size and the order parameter becomes nontrivial, and it is a priori unclear which of the phenomena described above can be observed in the evolution of the photoemission spectrum. To clarify this point, we analyze the relation between the gap and the order parameter in the present case. In Fig.~\ref{Fig.:RhoVsDelta} we plot the gap size $\Delta$ (obtained from the low energy maximum of the photoemission spectra at the momentum where the gap closes) as a function of the  order parameter $|\rho_{12}|$, both in equilibrium, and in nonequilibrium for various excitation amplitudes and times after the pulse. Interestingly, although the photoexcited states are clearly nonthermal (and in particular have a large number of excess carriers with respect to the thermal state), the nonequilibrium ratio between the two quantities agrees with the equilibrium ratio for different temperatures.

The strong correlation between the order parameter and the gap indicates that one should be able to see some characteristics of the nonthermal critical behavior, which was analyzed in terms of the  the order parameter $\rho$ (Fig.~\ref{Fig.:Relax_time}) also in the photoemission spectrum. One indicator for the existence of a nonthermal critical point is the two-time relaxation process displayed in Fig.~\ref{Fig.:Relax_time}(a). In the current setup this behavior is however difficult to extract from the spectrum due to several reasons: The crossover to the thermalization happens when the order parameter is already very small, and thus the resulting differences in the spectrum are minor. Furthermore, in a realistic experimental setup the longer time dynamics strongly depends on the coupling to  additional degrees of freedom, such as lattice-modulation phonons, which are not included in our description.  However, the more direct indication of the nonthermal critical point, which is the slowdown of the dynamics at the nonthermal threshold, is clearly visible in the photoemission spectrum: In Fig.~\ref{Fig.:Relax_time}(b) we plot the relaxation time $\tau_I$ which is extracted from an exponential fit to the relaxation of the local part of the photoemission intensity at the Fermi energy $I_{loc}(\omega=0)$. It displays a slowdown in the same region as order parameter. The analysis of the timescales in the  photoemission spectra therefore gives an experimentally accessible test for the existence of a nonthermal critical point. 

The almost universal ratio between the order parameter and the gap in Fig.~\ref{Fig.:RhoVsDelta} can also be used as a phenomenological way to analyze the question to what extent the change of the interaction due to screening plays a role in the dynamics of the excitonic condensate. This question is difficult to answer in general, as there is not a single matrix element (or single mode) and energy range of the screened interaction $W(\omega)$ which would have a dominant contribution, while the photo-induced changes of $W$ are partly opposite in various energy ranges as discussed above. However, motivated by a mean-field description where the gap is proportional to the order parameter and the ratio is set by the interaction, one can simply take the ratio of the gap and order parameter as a measure for the relevant pairing interaction. For the different excitations displayed in Fig.~\ref{Fig.:RhoVsDelta} the photoexcited  carrier numbers are very different, but nevertheless the ratio of gap and order parameter is comparable (compare, e.g., the  
data close to the point $\Delta=0.5$ and  $\rho=0.21$).
Furthermore, this ratio is also very similar to the ratio measured in the simple mean-field calculation. This can be interpreted in the sense that
the screening effects are not dominant for the melting of the condensate, but the main mechanism is the breaking of  excitons by photo-induced carriers through impact ionization. 

\section{Conclusion}\label{Sec.:Conclusion}

We have used the non-equilibrium generalization of the self-consistent GW method to simulate and analyze the dynamics of a pulse-excited excitonic insulator. We compared the results with the Hartree-Fock theory in equilibrium, where basic features of the spectrum agreed on the qualitative level, while the lack of memory effects and momentum transfer implies the absence of thermalization.  The latter effects are captured by the GW method, which allows us to discuss the interplay of excitonic formation, impact-ionization, thermalization and dynamical screening. The method should yield qualitatively accurate results in the limit of weak electron-electron coupling. 

We found that photo-excitation leads to a melting of the excitonic order parameter. Depending on the pulse strength one can eiter induce a transition into either a non-thermal trapped state 
with reduced condensate density
or a melting of the long-range order. A slowdown of the dynamics is observed at the threshold which separates the two regimes. 
The threshold field for fast melting of the order parameter turns out not to be related to the thermal critical point, which indicates that the two relaxation regimes are separated by a nonthermal critical point.\cite{Berges2008,tsuji2013} Consequently, we find that the order parameter exhibits an exponential dynamics with two characteristic time scales: The first is attributed to the dephasing dynamics with a critical slowdown at the nonthermal critical point. 
The second characteristic timescale is the time of thermalization which exhibits a slowdown as one approaches the thermal critical point. The whole dynamics strongly resembles the dynamics of the antiferromagnet in the weakly interacting Hubbard model,\cite{tsuji2013} where a transition from the antiferromagnet to paramagnet has been studied in an interaction-quench setup. 

The mechanism for melting the gap is a transfer of kinetic energy to the condensate, which also leads to impact ionization, and a stronger increase of the carrier number with absorbed energy below the threshold. In the present model, the modifications in the dynamically screened interaction by photo-excitation seems to be a secondary mechanism for the melting of the condensate: A direct analysis of the interaction shows significant modifications after photodoping, but the effect is opposite at low frequences and in the energy range of the gap, making the overall consequences subtle. On a more phenomenological level, one finds that the ratio of the order parameter and the gap (as measured in photoemission spectroscopy), which may be taken as an overall measure for the interaction, is almost independent of the number of photoexcited carriers, and remains comparable to the value obtained with a static interaction even for strong photoexcitation. Concerning real materials with more than one band, the quantitative effect of screening may be different, but one can expect that incoherent scattering and impact ionization remains a universal mechanism for the melting of the condensate. Such systems may therefore be a good candidate to eventually see a nonthermal critical point in an experiment. 

Furthermore, as shown in recent experiments \cite{dalConte2012,mohr2011,porer2014} the ultrafast response of correlated materials allows us to disentangle different degrees of freedom in real time. The example of TiSe$_2$ showed\cite{porer2014} that after a short pulse the system is trapped in a non-thermal CDW state without excitonic correlations. The possibility of disentagling the intertwined orders by ultrafast dynamics could be applicable to a broad variety of materials. By combining the method introduced in this work with the nonequilibrium version of the self-consistent Migdal approximation, see for instance Ref.~\onlinecite{murakami2015}, one can theoretically investigate the important question of the interplay between the electronic and structural contribution to the CDW. Another possible extension is the combination with the extended dynamical mean field theory, which allows us to treat stronger electron-electron interactions.\cite{sun2002,ayral2013,golez2015}

\acknowledgements

We would like to thank H.~Strand, L.~Boehnke and C.~Monney for helpful discussions.
The  calculations were run on the supercomputer HLRN of the North-German 
Supercomputing Alliance, and on the Beo04 cluster at the University of Fribourg. 
DG and PW acknowledge support from FP7 ERC starting grant No. 278023 and from SNSF Grant No. 200021\_140648,
ME acknowledges support from by the Deutsche Forschungsgemeinschaft within the Sonderforschungsbereich 925
(project B4).
\bibliography{BibTex/tdmft,BibTex/Books,BibTex/Polarons}

\begin{thebibliography}{41}
\expandafter\ifx\csname natexlab\endcsname\relax\def\natexlab#1{#1}\fi
\expandafter\ifx\csname bibnamefont\endcsname\relax
  \def\bibnamefont#1{#1}\fi
\expandafter\ifx\csname bibfnamefont\endcsname\relax
  \def\bibfnamefont#1{#1}\fi
\expandafter\ifx\csname citenamefont\endcsname\relax
  \def\citenamefont#1{#1}\fi
\expandafter\ifx\csname url\endcsname\relax
  \def\url#1{\texttt{#1}}\fi
\expandafter\ifx\csname urlprefix\endcsname\relax\def\urlprefix{URL }\fi
\providecommand{\bibinfo}[2]{#2}
\providecommand{\eprint}[2][]{\url{#2}}

\bibitem[{\citenamefont{{Giannetti} et~al.}(2016)\citenamefont{{Giannetti},
  {Capone}, {Fausti}, {Fabrizio}, {Parmigiani}, and
  {Mihailovic}}}]{giannetti2016}
\bibinfo{author}{\bibfnamefont{C.}~\bibnamefont{{Giannetti}}},
  \bibinfo{author}{\bibfnamefont{M.}~\bibnamefont{{Capone}}},
  \bibinfo{author}{\bibfnamefont{D.}~\bibnamefont{{Fausti}}},
  \bibinfo{author}{\bibfnamefont{M.}~\bibnamefont{{Fabrizio}}},
  \bibinfo{author}{\bibfnamefont{F.}~\bibnamefont{{Parmigiani}}},
  \bibnamefont{and}
  \bibinfo{author}{\bibfnamefont{D.}~\bibnamefont{{Mihailovic}}},
  \bibinfo{journal}{ArXiv e-prints}  (\bibinfo{year}{2016}),
  \eprint{1601.07204}.

\bibitem[{\citenamefont{Fausti et~al.}(2011)\citenamefont{Fausti, Tobey, Dean,
  Kaiser, Dienst, Hoffmann, Pyon, Takayama, Takagi, and
  Cavalleri}}]{fausti2011}
\bibinfo{author}{\bibfnamefont{D.}~\bibnamefont{Fausti}},
  \bibinfo{author}{\bibfnamefont{R.~I.} \bibnamefont{Tobey}},
  \bibinfo{author}{\bibfnamefont{N.}~\bibnamefont{Dean}},
  \bibinfo{author}{\bibfnamefont{S.}~\bibnamefont{Kaiser}},
  \bibinfo{author}{\bibfnamefont{A.}~\bibnamefont{Dienst}},
  \bibinfo{author}{\bibfnamefont{M.~C.} \bibnamefont{Hoffmann}},
  \bibinfo{author}{\bibfnamefont{S.}~\bibnamefont{Pyon}},
  \bibinfo{author}{\bibfnamefont{T.}~\bibnamefont{Takayama}},
  \bibinfo{author}{\bibfnamefont{H.}~\bibnamefont{Takagi}}, \bibnamefont{and}
  \bibinfo{author}{\bibfnamefont{A.}~\bibnamefont{Cavalleri}},
  \bibinfo{journal}{Science} \textbf{\bibinfo{volume}{331}},
  \bibinfo{pages}{189} (\bibinfo{year}{2011}),
  \eprint{http://www.sciencemag.org/content/331/6014/189.full.pdf},
  \urlprefix\url{http://www.sciencemag.org/content/331/6014/189.abstract}.

\bibitem[{\citenamefont{Mott}(1961)}]{mott1961}
\bibinfo{author}{\bibfnamefont{N.~F.} \bibnamefont{Mott}},
  \bibinfo{journal}{Philosophical Magazine} \textbf{\bibinfo{volume}{6}},
  \bibinfo{pages}{287} (\bibinfo{year}{1961}),
  \eprint{http://dx.doi.org/10.1080/14786436108243318},
  \urlprefix\url{http://dx.doi.org/10.1080/14786436108243318}.

\bibitem[{\citenamefont{Knox}(1963)}]{knox1963}
\bibinfo{author}{\bibfnamefont{R.}~\bibnamefont{Knox}},
  \bibinfo{journal}{Academic, New York}  (\bibinfo{year}{1963}).

\bibitem[{\citenamefont{J\'erome et~al.}(1967)\citenamefont{J\'erome, Rice, and
  Kohn}}]{jerome1967}
\bibinfo{author}{\bibfnamefont{D.}~\bibnamefont{J\'erome}},
  \bibinfo{author}{\bibfnamefont{T.~M.} \bibnamefont{Rice}}, \bibnamefont{and}
  \bibinfo{author}{\bibfnamefont{W.}~\bibnamefont{Kohn}},
  \bibinfo{journal}{Phys. Rev.} \textbf{\bibinfo{volume}{158}},
  \bibinfo{pages}{462} (\bibinfo{year}{1967}),
  \urlprefix\url{http://link.aps.org/doi/10.1103/PhysRev.158.462}.

\bibitem[{\citenamefont{Kohn}(1967)}]{kohn1967}
\bibinfo{author}{\bibfnamefont{W.}~\bibnamefont{Kohn}}, \bibinfo{journal}{Phys.
  Rev. Lett.} \textbf{\bibinfo{volume}{19}}, \bibinfo{pages}{439}
  (\bibinfo{year}{1967}),
  \urlprefix\url{http://link.aps.org/doi/10.1103/PhysRevLett.19.439}.

\bibitem[{\citenamefont{Di~Salvo et~al.}(1976)\citenamefont{Di~Salvo, Moncton,
  and Waszczak}}]{disalvo2976}
\bibinfo{author}{\bibfnamefont{F.~J.} \bibnamefont{Di~Salvo}},
  \bibinfo{author}{\bibfnamefont{D.~E.} \bibnamefont{Moncton}},
  \bibnamefont{and} \bibinfo{author}{\bibfnamefont{J.~V.}
  \bibnamefont{Waszczak}}, \bibinfo{journal}{Phys. Rev. B}
  \textbf{\bibinfo{volume}{14}}, \bibinfo{pages}{4321} (\bibinfo{year}{1976}),
  \urlprefix\url{http://link.aps.org/doi/10.1103/PhysRevB.14.4321}.

\bibitem[{\citenamefont{Kidd et~al.}(2002)\citenamefont{Kidd, Miller, Chou, and
  Chiang}}]{kidd2002}
\bibinfo{author}{\bibfnamefont{T.~E.} \bibnamefont{Kidd}},
  \bibinfo{author}{\bibfnamefont{T.}~\bibnamefont{Miller}},
  \bibinfo{author}{\bibfnamefont{M.~Y.} \bibnamefont{Chou}}, \bibnamefont{and}
  \bibinfo{author}{\bibfnamefont{T.-C.} \bibnamefont{Chiang}},
  \bibinfo{journal}{Phys. Rev. Lett.} \textbf{\bibinfo{volume}{88}},
  \bibinfo{pages}{226402} (\bibinfo{year}{2002}),
  \urlprefix\url{http://link.aps.org/doi/10.1103/PhysRevLett.88.226402}.

\bibitem[{\citenamefont{Rossnagel et~al.}(2002)\citenamefont{Rossnagel, Kipp,
  and Skibowski}}]{rossnagel2002}
\bibinfo{author}{\bibfnamefont{K.}~\bibnamefont{Rossnagel}},
  \bibinfo{author}{\bibfnamefont{L.}~\bibnamefont{Kipp}}, \bibnamefont{and}
  \bibinfo{author}{\bibfnamefont{M.}~\bibnamefont{Skibowski}},
  \bibinfo{journal}{Phys. Rev. B} \textbf{\bibinfo{volume}{65}},
  \bibinfo{pages}{235101} (\bibinfo{year}{2002}),
  \urlprefix\url{http://link.aps.org/doi/10.1103/PhysRevB.65.235101}.

\bibitem[{\citenamefont{Monney et~al.}(2009)\citenamefont{Monney, Cercellier,
  Clerc, Battaglia, Schwier, Didiot, Garnier, Beck, Aebi, Berger
  et~al.}}]{monney2009}
\bibinfo{author}{\bibfnamefont{C.}~\bibnamefont{Monney}},
  \bibinfo{author}{\bibfnamefont{H.}~\bibnamefont{Cercellier}},
  \bibinfo{author}{\bibfnamefont{F.}~\bibnamefont{Clerc}},
  \bibinfo{author}{\bibfnamefont{C.}~\bibnamefont{Battaglia}},
  \bibinfo{author}{\bibfnamefont{E.~F.} \bibnamefont{Schwier}},
  \bibinfo{author}{\bibfnamefont{C.}~\bibnamefont{Didiot}},
  \bibinfo{author}{\bibfnamefont{M.~G.} \bibnamefont{Garnier}},
  \bibinfo{author}{\bibfnamefont{H.}~\bibnamefont{Beck}},
  \bibinfo{author}{\bibfnamefont{P.}~\bibnamefont{Aebi}},
  \bibinfo{author}{\bibfnamefont{H.}~\bibnamefont{Berger}},
  \bibnamefont{et~al.}, \bibinfo{journal}{Phys. Rev. B}
  \textbf{\bibinfo{volume}{79}}, \bibinfo{pages}{045116}
  (\bibinfo{year}{2009}),
  \urlprefix\url{http://link.aps.org/doi/10.1103/PhysRevB.79.045116}.

\bibitem[{\citenamefont{Pillo et~al.}(2000)\citenamefont{Pillo, Hayoz, Berger,
  L\'evy, Schlapbach, and Aebi}}]{pillo2000}
\bibinfo{author}{\bibfnamefont{T.}~\bibnamefont{Pillo}},
  \bibinfo{author}{\bibfnamefont{J.}~\bibnamefont{Hayoz}},
  \bibinfo{author}{\bibfnamefont{H.}~\bibnamefont{Berger}},
  \bibinfo{author}{\bibfnamefont{F.}~\bibnamefont{L\'evy}},
  \bibinfo{author}{\bibfnamefont{L.}~\bibnamefont{Schlapbach}},
  \bibnamefont{and} \bibinfo{author}{\bibfnamefont{P.}~\bibnamefont{Aebi}},
  \bibinfo{journal}{Phys. Rev. B} \textbf{\bibinfo{volume}{61}},
  \bibinfo{pages}{16213} (\bibinfo{year}{2000}),
  \urlprefix\url{http://link.aps.org/doi/10.1103/PhysRevB.61.16213}.

\bibitem[{\citenamefont{Hughes}(1977)}]{hughes1977}
\bibinfo{author}{\bibfnamefont{H.~P.} \bibnamefont{Hughes}},
  \bibinfo{journal}{Journal of Physics C: Solid State Physics}
  \textbf{\bibinfo{volume}{10}}, \bibinfo{pages}{L319} (\bibinfo{year}{1977}),
  \urlprefix\url{http://stacks.iop.org/0022-3719/10/i=11/a=009}.

\bibitem[{\citenamefont{Kaneko et~al.}(2015)\citenamefont{Kaneko, Zenker,
  Fehske, and Ohta}}]{kaneko2015}
\bibinfo{author}{\bibfnamefont{T.}~\bibnamefont{Kaneko}},
  \bibinfo{author}{\bibfnamefont{B.}~\bibnamefont{Zenker}},
  \bibinfo{author}{\bibfnamefont{H.}~\bibnamefont{Fehske}}, \bibnamefont{and}
  \bibinfo{author}{\bibfnamefont{Y.}~\bibnamefont{Ohta}},
  \bibinfo{journal}{Phys. Rev. B} \textbf{\bibinfo{volume}{92}},
  \bibinfo{pages}{115106} (\bibinfo{year}{2015}),
  \urlprefix\url{http://link.aps.org/doi/10.1103/PhysRevB.92.115106}.

\bibitem[{\citenamefont{Kaneko et~al.}(2013)\citenamefont{Kaneko, Toriyama,
  Konishi, and Ohta}}]{kaneko2013}
\bibinfo{author}{\bibfnamefont{T.}~\bibnamefont{Kaneko}},
  \bibinfo{author}{\bibfnamefont{T.}~\bibnamefont{Toriyama}},
  \bibinfo{author}{\bibfnamefont{T.}~\bibnamefont{Konishi}}, \bibnamefont{and}
  \bibinfo{author}{\bibfnamefont{Y.}~\bibnamefont{Ohta}},
  \bibinfo{journal}{Phys. Rev. B} \textbf{\bibinfo{volume}{87}},
  \bibinfo{pages}{035121} (\bibinfo{year}{2013}),
  \urlprefix\url{http://link.aps.org/doi/10.1103/PhysRevB.87.035121}.

\bibitem[{\citenamefont{Zenker et~al.}(2014)\citenamefont{Zenker, Fehske, and
  Beck}}]{zenker2014}
\bibinfo{author}{\bibfnamefont{B.}~\bibnamefont{Zenker}},
  \bibinfo{author}{\bibfnamefont{H.}~\bibnamefont{Fehske}}, \bibnamefont{and}
  \bibinfo{author}{\bibfnamefont{H.}~\bibnamefont{Beck}},
  \bibinfo{journal}{Phys. Rev. B} \textbf{\bibinfo{volume}{90}},
  \bibinfo{pages}{195118} (\bibinfo{year}{2014}),
  \urlprefix\url{http://link.aps.org/doi/10.1103/PhysRevB.90.195118}.

\bibitem[{\citenamefont{Zocher et~al.}(2011)\citenamefont{Zocher, Timm, and
  Brydon}}]{zocher2011}
\bibinfo{author}{\bibfnamefont{B.}~\bibnamefont{Zocher}},
  \bibinfo{author}{\bibfnamefont{C.}~\bibnamefont{Timm}}, \bibnamefont{and}
  \bibinfo{author}{\bibfnamefont{P.~M.~R.} \bibnamefont{Brydon}},
  \bibinfo{journal}{Phys. Rev. B} \textbf{\bibinfo{volume}{84}},
  \bibinfo{pages}{144425} (\bibinfo{year}{2011}),
  \urlprefix\url{http://link.aps.org/doi/10.1103/PhysRevB.84.144425}.

\bibitem[{\citenamefont{Kaneko et~al.}(2012)\citenamefont{Kaneko, Seki, and
  Ohta}}]{kaneko2012}
\bibinfo{author}{\bibfnamefont{T.}~\bibnamefont{Kaneko}},
  \bibinfo{author}{\bibfnamefont{K.}~\bibnamefont{Seki}}, \bibnamefont{and}
  \bibinfo{author}{\bibfnamefont{Y.}~\bibnamefont{Ohta}},
  \bibinfo{journal}{Phys. Rev. B} \textbf{\bibinfo{volume}{85}},
  \bibinfo{pages}{165135} (\bibinfo{year}{2012}),
  \urlprefix\url{http://link.aps.org/doi/10.1103/PhysRevB.85.165135}.

\bibitem[{\citenamefont{M\"ohr-Vorobeva
  et~al.}(2011)\citenamefont{M\"ohr-Vorobeva, Johnson, Beaud, Staub, De~Souza,
  Milne, Ingold, Demsar, Schaefer, and Titov}}]{mohr2011}
\bibinfo{author}{\bibfnamefont{E.}~\bibnamefont{M\"ohr-Vorobeva}},
  \bibinfo{author}{\bibfnamefont{S.~L.} \bibnamefont{Johnson}},
  \bibinfo{author}{\bibfnamefont{P.}~\bibnamefont{Beaud}},
  \bibinfo{author}{\bibfnamefont{U.}~\bibnamefont{Staub}},
  \bibinfo{author}{\bibfnamefont{R.}~\bibnamefont{De~Souza}},
  \bibinfo{author}{\bibfnamefont{C.}~\bibnamefont{Milne}},
  \bibinfo{author}{\bibfnamefont{G.}~\bibnamefont{Ingold}},
  \bibinfo{author}{\bibfnamefont{J.}~\bibnamefont{Demsar}},
  \bibinfo{author}{\bibfnamefont{H.}~\bibnamefont{Schaefer}}, \bibnamefont{and}
  \bibinfo{author}{\bibfnamefont{A.}~\bibnamefont{Titov}},
  \bibinfo{journal}{Phys. Rev. Lett.} \textbf{\bibinfo{volume}{107}},
  \bibinfo{pages}{036403} (\bibinfo{year}{2011}),
  \urlprefix\url{http://link.aps.org/doi/10.1103/PhysRevLett.107.036403}.

\bibitem[{\citenamefont{Rohwer et~al.}(2011)\citenamefont{Rohwer, Hellmann,
  Wiesenmayer, Sohrt, Stange, Slomski, Carr, Liu, Avila, Kall{\"a}ne
  et~al.}}]{rohwer2011}
\bibinfo{author}{\bibfnamefont{T.}~\bibnamefont{Rohwer}},
  \bibinfo{author}{\bibfnamefont{S.}~\bibnamefont{Hellmann}},
  \bibinfo{author}{\bibfnamefont{M.}~\bibnamefont{Wiesenmayer}},
  \bibinfo{author}{\bibfnamefont{C.}~\bibnamefont{Sohrt}},
  \bibinfo{author}{\bibfnamefont{A.}~\bibnamefont{Stange}},
  \bibinfo{author}{\bibfnamefont{B.}~\bibnamefont{Slomski}},
  \bibinfo{author}{\bibfnamefont{A.}~\bibnamefont{Carr}},
  \bibinfo{author}{\bibfnamefont{Y.}~\bibnamefont{Liu}},
  \bibinfo{author}{\bibfnamefont{L.~M.} \bibnamefont{Avila}},
  \bibinfo{author}{\bibfnamefont{M.}~\bibnamefont{Kall{\"a}ne}},
  \bibnamefont{et~al.}, \bibinfo{journal}{Nature}
  \textbf{\bibinfo{volume}{471}}, \bibinfo{pages}{490} (\bibinfo{year}{2011}).

\bibitem[{\citenamefont{Porer et~al.}(2014)\citenamefont{Porer, Leierseder,
  M{\'e}nard, Dachraoui, Mouchliadis, Perakis, Heinzmann, Demsar, Rossnagel,
  and Huber}}]{porer2014}
\bibinfo{author}{\bibfnamefont{M.}~\bibnamefont{Porer}},
  \bibinfo{author}{\bibfnamefont{U.}~\bibnamefont{Leierseder}},
  \bibinfo{author}{\bibfnamefont{J.-M.} \bibnamefont{M{\'e}nard}},
  \bibinfo{author}{\bibfnamefont{H.}~\bibnamefont{Dachraoui}},
  \bibinfo{author}{\bibfnamefont{L.}~\bibnamefont{Mouchliadis}},
  \bibinfo{author}{\bibfnamefont{I.}~\bibnamefont{Perakis}},
  \bibinfo{author}{\bibfnamefont{U.}~\bibnamefont{Heinzmann}},
  \bibinfo{author}{\bibfnamefont{J.}~\bibnamefont{Demsar}},
  \bibinfo{author}{\bibfnamefont{K.}~\bibnamefont{Rossnagel}},
  \bibnamefont{and} \bibinfo{author}{\bibfnamefont{R.}~\bibnamefont{Huber}},
  \bibinfo{journal}{Nature materials} \textbf{\bibinfo{volume}{13}},
  \bibinfo{pages}{857} (\bibinfo{year}{2014}).

\bibitem[{\citenamefont{Huber et~al.}(2001)\citenamefont{Huber, Tauser,
  Brodschelm, Bichler, Abstreiter, and Leitenstorfer}}]{huber2001}
\bibinfo{author}{\bibfnamefont{R.}~\bibnamefont{Huber}},
  \bibinfo{author}{\bibfnamefont{F.}~\bibnamefont{Tauser}},
  \bibinfo{author}{\bibfnamefont{A.}~\bibnamefont{Brodschelm}},
  \bibinfo{author}{\bibfnamefont{M.}~\bibnamefont{Bichler}},
  \bibinfo{author}{\bibfnamefont{G.}~\bibnamefont{Abstreiter}},
  \bibnamefont{and}
  \bibinfo{author}{\bibfnamefont{A.}~\bibnamefont{Leitenstorfer}},
  \bibinfo{journal}{Nature} \textbf{\bibinfo{volume}{414}},
  \bibinfo{pages}{286} (\bibinfo{year}{2001}).

\bibitem[{\citenamefont{Rossi and Kuhn}(2002)}]{rossi2002}
\bibinfo{author}{\bibfnamefont{F.}~\bibnamefont{Rossi}} \bibnamefont{and}
  \bibinfo{author}{\bibfnamefont{T.}~\bibnamefont{Kuhn}},
  \bibinfo{journal}{Rev. Mod. Phys.} \textbf{\bibinfo{volume}{74}},
  \bibinfo{pages}{895} (\bibinfo{year}{2002}),
  \urlprefix\url{http://link.aps.org/doi/10.1103/RevModPhys.74.895}.

\bibitem[{\citenamefont{El~Sayed et~al.}(1994)\citenamefont{El~Sayed, Schuster,
  Haug, Herzel, and Henneberger}}]{sayed1994}
\bibinfo{author}{\bibfnamefont{K.}~\bibnamefont{El~Sayed}},
  \bibinfo{author}{\bibfnamefont{S.}~\bibnamefont{Schuster}},
  \bibinfo{author}{\bibfnamefont{H.}~\bibnamefont{Haug}},
  \bibinfo{author}{\bibfnamefont{F.}~\bibnamefont{Herzel}}, \bibnamefont{and}
  \bibinfo{author}{\bibfnamefont{K.}~\bibnamefont{Henneberger}},
  \bibinfo{journal}{Phys. Rev. B} \textbf{\bibinfo{volume}{49}},
  \bibinfo{pages}{7337} (\bibinfo{year}{1994}),
  \urlprefix\url{http://link.aps.org/doi/10.1103/PhysRevB.49.7337}.

\bibitem[{\citenamefont{B\'anyai et~al.}(1998)\citenamefont{B\'anyai, Vu,
  Mieck, and Haug}}]{banyai1998}
\bibinfo{author}{\bibfnamefont{L.}~\bibnamefont{B\'anyai}},
  \bibinfo{author}{\bibfnamefont{Q.~T.} \bibnamefont{Vu}},
  \bibinfo{author}{\bibfnamefont{B.}~\bibnamefont{Mieck}}, \bibnamefont{and}
  \bibinfo{author}{\bibfnamefont{H.}~\bibnamefont{Haug}},
  \bibinfo{journal}{Phys. Rev. Lett.} \textbf{\bibinfo{volume}{81}},
  \bibinfo{pages}{882} (\bibinfo{year}{1998}),
  \urlprefix\url{http://link.aps.org/doi/10.1103/PhysRevLett.81.882}.

\bibitem[{\citenamefont{Barankov and Levitov}(2006)}]{barankov2006}
\bibinfo{author}{\bibfnamefont{R.~A.} \bibnamefont{Barankov}} \bibnamefont{and}
  \bibinfo{author}{\bibfnamefont{L.~S.} \bibnamefont{Levitov}},
  \bibinfo{journal}{Phys. Rev. Lett.} \textbf{\bibinfo{volume}{96}},
  \bibinfo{pages}{230403} (\bibinfo{year}{2006}),
  \urlprefix\url{http://link.aps.org/doi/10.1103/PhysRevLett.96.230403}.

\bibitem[{\citenamefont{Sciolla and Biroli}(2013)}]{sciolla2013}
\bibinfo{author}{\bibfnamefont{B.}~\bibnamefont{Sciolla}} \bibnamefont{and}
  \bibinfo{author}{\bibfnamefont{G.}~\bibnamefont{Biroli}},
  \bibinfo{journal}{Phys. Rev. B} \textbf{\bibinfo{volume}{88}},
  \bibinfo{pages}{201110} (\bibinfo{year}{2013}),
  \urlprefix\url{http://link.aps.org/doi/10.1103/PhysRevB.88.201110}.

\bibitem[{\citenamefont{Tsuji et~al.}(2013)\citenamefont{Tsuji, Eckstein, and
  Werner}}]{tsuji2013}
\bibinfo{author}{\bibfnamefont{N.}~\bibnamefont{Tsuji}},
  \bibinfo{author}{\bibfnamefont{M.}~\bibnamefont{Eckstein}}, \bibnamefont{and}
  \bibinfo{author}{\bibfnamefont{P.}~\bibnamefont{Werner}},
  \bibinfo{journal}{Phys. Rev. Lett.} \textbf{\bibinfo{volume}{110}},
  \bibinfo{pages}{136404} (\bibinfo{year}{2013}),
  \urlprefix\url{http://link.aps.org/doi/10.1103/PhysRevLett.110.136404}.

\bibitem[{\citenamefont{Werner et~al.}(2012)\citenamefont{Werner, Tsuji, and
  Eckstein}}]{werner2012}
\bibinfo{author}{\bibfnamefont{P.}~\bibnamefont{Werner}},
  \bibinfo{author}{\bibfnamefont{N.}~\bibnamefont{Tsuji}}, \bibnamefont{and}
  \bibinfo{author}{\bibfnamefont{M.}~\bibnamefont{Eckstein}},
  \bibinfo{journal}{Phys. Rev. B} \textbf{\bibinfo{volume}{86}},
  \bibinfo{pages}{205101} (\bibinfo{year}{2012}),
  \urlprefix\url{http://link.aps.org/doi/10.1103/PhysRevB.86.205101}.

\bibitem[{\citenamefont{Berges et~al.}(2008)\citenamefont{Berges, Rothkopf, and
  Schmidt}}]{Berges2008}
\bibinfo{author}{\bibfnamefont{J.}~\bibnamefont{Berges}},
  \bibinfo{author}{\bibfnamefont{A.}~\bibnamefont{Rothkopf}}, \bibnamefont{and}
  \bibinfo{author}{\bibfnamefont{J.}~\bibnamefont{Schmidt}},
  \bibinfo{journal}{Phys. Rev. Lett.} \textbf{\bibinfo{volume}{101}},
  \bibinfo{pages}{041603} (\bibinfo{year}{2008}),
  \urlprefix\url{http://link.aps.org/doi/10.1103/PhysRevLett.101.041603}.

\bibitem[{\citenamefont{Nowak et~al.}(2011)\citenamefont{Nowak, Sexty, and
  Gasenzer}}]{nowak2011}
\bibinfo{author}{\bibfnamefont{B.}~\bibnamefont{Nowak}},
  \bibinfo{author}{\bibfnamefont{D.}~\bibnamefont{Sexty}}, \bibnamefont{and}
  \bibinfo{author}{\bibfnamefont{T.}~\bibnamefont{Gasenzer}},
  \bibinfo{journal}{Phys. Rev. B} \textbf{\bibinfo{volume}{84}},
  \bibinfo{pages}{020506} (\bibinfo{year}{2011}),
  \urlprefix\url{http://link.aps.org/doi/10.1103/PhysRevB.84.020506}.

\bibitem[{\citenamefont{Stefanucci and van Leeuwen}(2013)}]{stefanucci2013}
\bibinfo{author}{\bibfnamefont{G.}~\bibnamefont{Stefanucci}} \bibnamefont{and}
  \bibinfo{author}{\bibfnamefont{R.}~\bibnamefont{van Leeuwen}},
  \emph{\bibinfo{title}{Nonequilibrium Many-Body Theory of Quantum Systems: A
  Modern Introduction}} (\bibinfo{publisher}{Cambridge University Press},
  \bibinfo{year}{2013}).

\bibitem[{\citenamefont{Bronold and Fehske}(2006)}]{bronold2006}
\bibinfo{author}{\bibfnamefont{F.~X.} \bibnamefont{Bronold}} \bibnamefont{and}
  \bibinfo{author}{\bibfnamefont{H.}~\bibnamefont{Fehske}},
  \bibinfo{journal}{Phys. Rev. B} \textbf{\bibinfo{volume}{74}},
  \bibinfo{pages}{165107} (\bibinfo{year}{2006}),
  \urlprefix\url{http://link.aps.org/doi/10.1103/PhysRevB.74.165107}.

\bibitem[{\citenamefont{Kaneko and Ohta}(2014)}]{kaneko2014}
\bibinfo{author}{\bibfnamefont{T.}~\bibnamefont{Kaneko}} \bibnamefont{and}
  \bibinfo{author}{\bibfnamefont{Y.}~\bibnamefont{Ohta}},
  \bibinfo{journal}{Phys. Rev. B} \textbf{\bibinfo{volume}{90}},
  \bibinfo{pages}{245144} (\bibinfo{year}{2014}),
  \urlprefix\url{http://link.aps.org/doi/10.1103/PhysRevB.90.245144}.

\bibitem[{\citenamefont{Aoki et~al.}(2014)\citenamefont{Aoki, Tsuji, Eckstein,
  Kollar, Oka, and Werner}}]{aoki2014_rev}
\bibinfo{author}{\bibfnamefont{H.}~\bibnamefont{Aoki}},
  \bibinfo{author}{\bibfnamefont{N.}~\bibnamefont{Tsuji}},
  \bibinfo{author}{\bibfnamefont{M.}~\bibnamefont{Eckstein}},
  \bibinfo{author}{\bibfnamefont{M.}~\bibnamefont{Kollar}},
  \bibinfo{author}{\bibfnamefont{T.}~\bibnamefont{Oka}}, \bibnamefont{and}
  \bibinfo{author}{\bibfnamefont{P.}~\bibnamefont{Werner}},
  \bibinfo{journal}{Rev. Mod. Phys.} \textbf{\bibinfo{volume}{86}},
  \bibinfo{pages}{779} (\bibinfo{year}{2014}),
  \urlprefix\url{http://link.aps.org/doi/10.1103/RevModPhys.86.779}.

\bibitem[{\citenamefont{Freericks et~al.}(2009)\citenamefont{Freericks,
  Krishnamurthy, and Pruschke}}]{freericks09}
\bibinfo{author}{\bibfnamefont{J.~K.} \bibnamefont{Freericks}},
  \bibinfo{author}{\bibfnamefont{H.~R.} \bibnamefont{Krishnamurthy}},
  \bibnamefont{and} \bibinfo{author}{\bibfnamefont{T.}~\bibnamefont{Pruschke}},
  \bibinfo{journal}{Phys. Rev. Lett.} \textbf{\bibinfo{volume}{102}},
  \bibinfo{pages}{136401} (\bibinfo{year}{2009}).

\bibitem[{\citenamefont{Werner et~al.}(2014)\citenamefont{Werner, Held, and
  Eckstein}}]{werner2014}
\bibinfo{author}{\bibfnamefont{P.}~\bibnamefont{Werner}},
  \bibinfo{author}{\bibfnamefont{K.}~\bibnamefont{Held}}, \bibnamefont{and}
  \bibinfo{author}{\bibfnamefont{M.}~\bibnamefont{Eckstein}},
  \bibinfo{journal}{Phys. Rev. B} \textbf{\bibinfo{volume}{90}},
  \bibinfo{pages}{235102} (\bibinfo{year}{2014}),
  \urlprefix\url{http://link.aps.org/doi/10.1103/PhysRevB.90.235102}.

\bibitem[{\citenamefont{Dal~Conte et~al.}(2012)\citenamefont{Dal~Conte,
  Giannetti, Coslovich, Cilento, Bossini, Abebaw, Banfi, Ferrini, Eisaki,
  Greven et~al.}}]{dalConte2012}
\bibinfo{author}{\bibfnamefont{S.}~\bibnamefont{Dal~Conte}},
  \bibinfo{author}{\bibfnamefont{C.}~\bibnamefont{Giannetti}},
  \bibinfo{author}{\bibfnamefont{G.}~\bibnamefont{Coslovich}},
  \bibinfo{author}{\bibfnamefont{F.}~\bibnamefont{Cilento}},
  \bibinfo{author}{\bibfnamefont{D.}~\bibnamefont{Bossini}},
  \bibinfo{author}{\bibfnamefont{T.}~\bibnamefont{Abebaw}},
  \bibinfo{author}{\bibfnamefont{F.}~\bibnamefont{Banfi}},
  \bibinfo{author}{\bibfnamefont{G.}~\bibnamefont{Ferrini}},
  \bibinfo{author}{\bibfnamefont{H.}~\bibnamefont{Eisaki}},
  \bibinfo{author}{\bibfnamefont{M.}~\bibnamefont{Greven}},
  \bibnamefont{et~al.}, \bibinfo{journal}{Science}
  \textbf{\bibinfo{volume}{335}}, \bibinfo{pages}{1600} (\bibinfo{year}{2012}),
  ISSN \bibinfo{issn}{0036-8075},
  \eprint{http://science.sciencemag.org/content/335/6076/1600.full.pdf},
  \urlprefix\url{http://science.sciencemag.org/content/335/6076/1600}.

\bibitem[{\citenamefont{Murakami et~al.}(2015)\citenamefont{Murakami, Werner,
  Tsuji, and Aoki}}]{murakami2015}
\bibinfo{author}{\bibfnamefont{Y.}~\bibnamefont{Murakami}},
  \bibinfo{author}{\bibfnamefont{P.}~\bibnamefont{Werner}},
  \bibinfo{author}{\bibfnamefont{N.}~\bibnamefont{Tsuji}}, \bibnamefont{and}
  \bibinfo{author}{\bibfnamefont{H.}~\bibnamefont{Aoki}},
  \bibinfo{journal}{Phys. Rev. B} \textbf{\bibinfo{volume}{91}},
  \bibinfo{pages}{045128} (\bibinfo{year}{2015}),
  \urlprefix\url{http://link.aps.org/doi/10.1103/PhysRevB.91.045128}.

\bibitem[{\citenamefont{Sun and Kotliar}(2002)}]{sun2002}
\bibinfo{author}{\bibfnamefont{P.}~\bibnamefont{Sun}} \bibnamefont{and}
  \bibinfo{author}{\bibfnamefont{G.}~\bibnamefont{Kotliar}},
  \bibinfo{journal}{Phys. Rev. B} \textbf{\bibinfo{volume}{66}},
  \bibinfo{pages}{085120} (\bibinfo{year}{2002}),
  \urlprefix\url{http://link.aps.org/doi/10.1103/PhysRevB.66.085120}.

\bibitem[{\citenamefont{Ayral et~al.}(2013)\citenamefont{Ayral, Biermann, and
  Werner}}]{ayral2013}
\bibinfo{author}{\bibfnamefont{T.}~\bibnamefont{Ayral}},
  \bibinfo{author}{\bibfnamefont{S.}~\bibnamefont{Biermann}}, \bibnamefont{and}
  \bibinfo{author}{\bibfnamefont{P.}~\bibnamefont{Werner}},
  \bibinfo{journal}{Phys. Rev. B} \textbf{\bibinfo{volume}{87}},
  \bibinfo{pages}{125149} (\bibinfo{year}{2013}),
  \urlprefix\url{http://link.aps.org/doi/10.1103/PhysRevB.87.125149}.

\bibitem[{\citenamefont{Gole\ifmmode~\check{z}\else \v{z}\fi{}
  et~al.}(2015)\citenamefont{Gole\ifmmode~\check{z}\else \v{z}\fi{}, Eckstein,
  and Werner}}]{golez2015}
\bibinfo{author}{\bibfnamefont{D.}~\bibnamefont{Gole\ifmmode~\check{z}\else
  \v{z}\fi{}}}, \bibinfo{author}{\bibfnamefont{M.}~\bibnamefont{Eckstein}},
  \bibnamefont{and} \bibinfo{author}{\bibfnamefont{P.}~\bibnamefont{Werner}},
  \bibinfo{journal}{Phys. Rev. B} \textbf{\bibinfo{volume}{92}},
  \bibinfo{pages}{195123} (\bibinfo{year}{2015}),
  \urlprefix\url{http://link.aps.org/doi/10.1103/PhysRevB.92.195123}.

\end{thebibliography}

\end{document}